\documentclass[
 reprint,
superscriptaddress,
 amsmath,amssymb,
 aps,
prd,
nofootinbib]{revtex4-2}

\usepackage{graphicx}
\usepackage{dcolumn}
\usepackage{bm}
\usepackage{hyperref}
\usepackage{cleveref}
\hypersetup{
citecolor=blue,
 colorlinks=true,
 linkcolor=blue,
 filecolor=magenta,
 urlcolor=cyan,
}

\newcommand{\apjl}{Astrophysical Journal, Letters}

\newcommand{\mnras}{Monthly Notices of the Royal Astronomical Society}

\usepackage{xcolor}
\usepackage{soul}
\usepackage[normalem]{ulem}
\usepackage{xspace}
\usepackage{booktabs}

\newcommand{\dyb}{\ensuremath{d^*}\xspace}
\newcommand{\dybnum}{\ensuremath{d^*(2380)}\xspace}
\newcommand{\gwd}{\ensuremath{g_{\omega d^*}}\xspace}
\newcommand{\gsd}{\ensuremath{g_{\sigma d^*}}\xspace}
\newcommand{\pdyb}{\ensuremath{P^0_{d^*}}\xspace}
\newcommand{\cs}{\ensuremath{\left <(c_s/c)^2 \right >}\xspace}

\begin{document}

\preprint{APS/123-QED}

\title{Exploring the role of \dyb hexaquarks on quark deconfinement and hybrid stars}


\author{Marcos O. Celi} \email{mceli@fcaglp.unlp.edu.ar}
\affiliation{Grupo de Astrof\'isica de Remanentes Compactos,\\ Facultad de Ciencias Astron{\'o}micas y
  Geof{\'i}sicas, Universidad Nacional de La Plata,\\ Paseo del Bosque
  S/N, La Plata (1900), Argentina.}
\affiliation{CONICET, Godoy Cruz 2290, Buenos Aires (1425), Argentina.}
\affiliation{Universidade Federal do ABC, Centro de Ciências Naturais e Humanas, Avenida dos
Estados 5001-Bangú, CEP 09210-580, Santo André, SP, Brazil}

\author{Mauro Mariani} \email{mmariani@fcaglp.unlp.edu.ar}
\affiliation{Grupo de Astrof\'isica de Remanentes Compactos,\\ Facultad de Ciencias Astron{\'o}micas y
  Geof{\'i}sicas, Universidad Nacional de La Plata,\\ Paseo del Bosque
  S/N, La Plata (1900), Argentina.}
\affiliation{CONICET, Godoy Cruz 2290, Buenos Aires (1425), Argentina.}

\author{Rajesh Kumar} 
\affiliation{Center for Nuclear Research, Department of Physics, Kent State University, Kent, OH 44243 USA}
\affiliation{Department of Physics, MRPD Government College, Talwara, Punjab 144216 India}

\author{Mikhail Bashkanov}
\affiliation{Department of Physics, University of York, , Heslington, York, Y010 5DD, UK}

\author{Milva G. Orsaria}
\affiliation{Grupo de Astrof\'isica de Remanentes Compactos,\\ Facultad de Ciencias Astron{\'o}micas y
  Geof{\'i}sicas, Universidad Nacional de La Plata,\\ Paseo del Bosque
  S/N, La Plata (1900), Argentina.}
\affiliation{CONICET, Godoy Cruz 2290, Buenos Aires (1425), Argentina.}
  
\author{Alessandro Pastore}
\affiliation{CEA, DES, IRESNE, DER, SPRC, F-13108 Saint Paul Lez Durance, France}

\author{Ignacio F. Ranea-Sandoval}
\affiliation{Grupo de Astrof\'isica de Remanentes Compactos,\\ Facultad de Ciencias Astron{\'o}micas y
  Geof{\'i}sicas, Universidad Nacional de La Plata,\\ Paseo del Bosque
  S/N, La Plata (1900), Argentina.}
\affiliation{CONICET, Godoy Cruz 2290, Buenos Aires (1425), Argentina.}

\author{Veronica Dexheimer}
\affiliation{Center for Nuclear Research, Department of Physics, Kent State University, Kent, OH 44243 USA}

\date{\today}

\begin{abstract}
We investigate the impact of the \dybnum hexaquark on the equation of state (EoS) of dense matter within hybrid stars (HSs) using the Chiral Mean-Field model (CMF). The hexaquark is included as a new degree of freedom in the hadronic phase, and its influence on the deconfinement transition to quark matter is explored. We re-parametrize the CMF model to ensure compatibility with recent astrophysical constraints, including the observation of massive pulsars and gravitational wave events. Our results show that the presence of \dyb significantly modifies the EoS, leading to a softening at high densities and a consequent reduction in the predicted maximum stellar masses. Furthermore, we examine the possibility of a first-order deconfinement phase transition within the context of the extended stability branch of slow stable HSs (SSHSs). We find that the presence of hexaquarks can delay the deconfinement phase transition and reduce the associated energy density gap, affecting the structure and stability of HSs. Our results suggest that, as the hexaquark appearance tends to destabilize stellar configurations, fine tuning of model parameters is required to obtain both the presence of hexaquarks and quark deconfinement in these systems.
In this scenario, the SSHS branch
plays a crucial role in obtaining HSs with hexaquarks that satisfy current astrophysical constraints.
Our work provides new insights into the role of exotic particles like \dyb in dense matter and the complex interplay between hadronic and quark degrees of freedom inside compact stellar objects.
\end{abstract}

\maketitle


\section{Introduction}
\label{sec:intro}

The research field of neutron star (NS) astrophysics is currently undergoing a revolution. With advancements in observational and computational capabilities, as well as the development of new theories and processing techniques, the high-energy astrophysical community now has access to results that were just a dream a few years ago. The detection of NS mergers by LIGO/Virgo, such as GW170817 \cite{Abbott:2017goo, Abbott:2017gwa, Abbott:2017moo} and GW190425 \cite{Abbott:2020goo}, along with NICER’s measurements of NS masses and radii, has provided unprecedented insights into ultra-dense matter. Observations of high-mass binary pulsars, including those with masses around $\sim 2 M_{\odot}$ (e.g., PSR J1614-2230 \cite{Demorest:2010ats, Arzoumanian:2018tny}, PSR J0348+0432 \cite{Antoniadis:2013amp}, and PSR J0740+6620 \cite{Cromartie:2020rsd, Fonseca:2021rma}), as well as NICER data from PSR J0030+0451 \cite{Riley:2019anv, Raaijmakers:2019anv, Miller:2019pjm}, PSR J0740+6620 \citep{Riley:2021anv, Miller:2021tro} (together with XMM-Newton data), and \cite{Salmi:2022tro, Dittmann:2024amp, Salmi:2024tro} (relying solely on NICER observations), PSR J0437-4715 \cite{Choudhury:2024anv, Reardon:2024tns}, and PSR J1231-1411 \cite{salmi:2024anv}, have set strong constraints on the EoS governing NS cores. In particular, despite PSR J0740+6620 being $\sim 50 \%$ more massive than PSR J0030+0451, the error bars in their radii are consistent with being nearly identical, challenging our understanding of NS interiors. Furthermore, analysis of GW170817 suggests NS radii between $9.1$–$13.3$ km \cite{yagi:2017aur}, while the absence of prompt collapse in its remnant implies a maximum NS mass of $\sim 2.3 M_{\odot}$ \cite{Shibata:2019cot}. GW190425 further supports that NSs with masses above $\sim 1.7 M_{\odot}$ are expected to have \mbox{radii $\gtrsim 11$ km} \cite{Abbott:2020goo}. These results challenge the determination of the EoS of NSs, and consequently our understanding of the matter inside these objects, particularly at densities several times the nuclear saturation density, $n_0$.

Moreover, there are relevant constraints from nuclear theory and laboratory experiments (see the review presented in Ref.~\cite{kumar:2024tae} and references therein). In particular, useful for our purposes are the constraints from chiral Effective Field Theory ($\chi$EFT) calculations, applicable in the range $n_0 \lesssim n \lesssim 2n_0$ \cite{Hebeler:2013eos, Lynn:2016ctn, Hu:2017nmp, Holt:2017eos, Drischler:2020hwd} (also see the work of \citet{Drischler:2021lma}, who used a Bayesian framework to better constrain the correlated truncation error associated with $\chi$EFT, resulting in a refined constraint for $\beta$-stable NS matter). Beyond this density range, it is necessary to use effective or phenomenological models. Among them, we recall the non-relativistic mean field models \cite{Vautherin:1972hfc,Potekhin:2013,Sharma:2015, Grasso:2019edf},  which are capable of describing only hadronic degrees of freedom, and relativistic mean field models (RMF) \cite{walecka:1974ato,Boguta:1977ddp,Typel:2018rmf,Malfatti:2020dba,Sedrakian:2023hbi} which can also account for additional degrees of freedom, such as hyperons (see Ref.\cite{Glendenning:2012} for more details).
Finally, to describe the possible presence of quarks within the interior of a NS, new families of models such as the chiral mean-field model (CMF) \cite{Dexheimer:2010ana,Kumar:2024mna,CruzCamacho:2024psi}, (non-)local Nambu-Jona-Lasinio-like models \cite{Blaschke:2005pto,Contrera:2010mpa,Contrera:2010npn,Malfatti:2017qhp}, and more recently those based on the Field Correlator Method \cite{Simonov:2007dtf,Simonov:2007vpt,Nefediev:2009daq,Mariani:2017ceh,Curin:2021hsw}) have been extensively used in the last few decades. In addition to these models, which involve a detailed microscopic description of matter, several parametric descriptions have become popular. Their main goal is to produce model-independent results. For example, we mention those presented in Refs.~\cite{Hebeler:2013eos,OBoyle:2020peo}, which are based on (generalized) piecewise polytropes used to describe the hadronic sector, and those based on the Constant Speed of Sound (CSS) parameterization for quarks and hadrons \cite{alford:2013gcf,tan:2020nse}. In addition, there are approaches entirely based on machine learning techniques, such as Gaussian Processes \cite{Keller:2023} or Neural Networks \cite{Morawski:2020}.

Within this scenario, several authors have analyzed the possibility of a first-order phase transition in the core of compact objects, where hadrons dissolve into quarks. The nature of this hadron-quark transition depends on the (unknown) surface tension at the interface. If the surface tension exceeds a critical threshold, the transition follows a Gibbs construction (e.g., Refs. \cite{Endo:2006cse,hempel:2013not,Wu:2019nse}); otherwise, it results in a Maxwell construction, which gives rise to a sharp phase transition \cite{Voskresensky:2002csa,Endo:2011roh,Wu:2018eoq}. A key feature of sharp transitions is the conversion speed between phases. \textit{Slow} conversions can produce stable hybrid stars (SSHS) even when \mbox{$\partial M/\partial \varepsilon _c < 0$} beyond the maximum mass configuration \cite{Pereira:2018pte,mariani:2019mhs,Malfatti:2020dba,Tonetto:2020dgm,rodriguez:2021hsw,Goncalves:2022ios,Mariani:2022omh,Ranea:2022bou,lugones:2023ama,Ranea:2023auq,Ranea:2023cmr,Rau:2023tfo,Rau:2023neo,Rather:2024roo}. Other studies focused on HSs with \textit{rapid} conversions \cite{Dexheimer:2010ana,Lenzi:2012hsi,Alvarez-Castillo:2019tfo}. Further details on hadron and quark EoS and their potential phase transitions in the cores of compact objects can be found in Refs. \cite{Baym:2018fht,Orsaria:2019pti,Lugones:2021pci}.

In the scope of HSs and beyond, several attempts have been made to present Lagrangians capable of describing both hadronic and quark phases in a unified manner \cite{Dexheimer:2008pna,steinheimer:2011hsup,bastian:2021auq,Bastian:2021pqh,Kumar:2024mna}. Among these, we focus on the CMF model \citep{Kumar:2024mna} which introduces an expansion into quark degrees of freedom, inspired by Polyakov Nambu Jona Lasinio (PNJL)-like models. In this context, the scalar field $\Phi$ acts as an order parameter associated with the PNJL-like effective potential, driving the transition from the confined to the deconfined phase. This approach enables the description of a crossover and a first-order phase transition (at low temperatures) using a unified Lagrangian for both phases.

Besides including hadrons and deconfined quarks, another possibility is the incorporation of different hexaquarks as new degrees of freedom in the description of dense matter. Hexaquarks constitute a large family of hypothetical particles made up of six quarks or antiquarks of any flavor, which can exist either as dibaryon molecules or as ``genuine multiquark" states. These particles have attracted significant interest because of their unique effects in dense matter. For example, in Ref.~\cite{Shahrbaf:2022sdi}, the authors explored the impact of including the hexaquark state with quark content $uuddss$ on the EoS of NSs, regardless of the lack of experimental confirmation for such a particle to date. Recent calculations of lattice quantum chromodynamics (Lattice-QCD) with nearly physical quark masses suggest that the $uuddss$ state is likely located in a closed vicinity of the $\Xi p$ threshold, indicating that this state is a spatially extended baryonic molecule, similar to the deuteron, rather than a compact genuine hexaquark~\cite{Sasaki:2018}. These findings are further supported by experimental femtoscopy observations at ALICE~\cite{Ohnishi:2021}. In this context,  the \dybnum hexaquark is thought to be a massive, positively charged non-strange particle with an integer spin ($J=3$). Moreover, it exhibits a $\Delta\Delta$ coupling so strong that some theoretical models treated it as a $70$~MeV bound state of two-$\Delta$'s, referred to as the \textit{Deltaron} \cite{Bashkanov:2019epo}. 

From a theoretical point of view, the microscopic description of the \dyb hexaquark remains challenging \cite{Dong:2023}. Several studies have examined the importance of \dybnum hexaquark for nuclear EoS in the context of NSs \cite{Bashkanov:2019epo,Vidana:2018tdi,Mantziris:2020nsm,Celi:2024doh}. However, since the early work by \citet{Faessler:1998},  limited progress has been made in describing this particle within RMF models. To address this, studies conducted by \citet{Mantziris:2020nsm} and \citet{Celi:2024doh} have revealed that, despite its substantial mass, the \dybnum hexaquark could exist at the same densities as $\Delta$ baryons or hyperons in the cores of NSs. As a result, its presence substantially modifies the behavior of other baryons when included in a given hadronic model. In these works, the possible parameter space characterizing the interaction coupling constants of \dyb hexaquark was narrowed down to ensure the compatibility of the studied EoSs with the available astrophysical constraints on NSs. This is particularly interesting because the \dybnum hexaquark is the first known non-trivial hexaquark with experimental evidence \cite{Adlarson:2011dcs}. Furthermore, \dyb can form a stable Bose-Einstein condensate \cite{Bashkanov:2020}, making it a possible candidate for dark matter \cite{Chan:2020,Beck:2021}. However, to date, there has been little or no experimental evidence concerning how such a hexaquark interacts with surrounding matter.

There is an ongoing program at A2@Mainz focused on measuring photoproduction processes on nuclear targets~\cite{Watts:2016}, ranging from ${}^{40}$Ca to ${}^{238}$U with the main goal of determining the neutron skin thickness and constraining nuclear EoS at saturation density. Besides measuring the neutron skin thickness~\cite{Tarbert:2014}, A2 collaboration is looking for the $d^*$ photoproduction inside nuclei, aiming to extract the $\dyb-N$ interaction strength. However, the analysis of such a process is very complicated and so far no results have been reported. That is why any astrophysically inspired constraint on the hexaquark-nucleon interaction strength could significantly simplify the analysis of these data and help eliminate certain backgrounds. Usually, experimental results in nuclear physics motivate the boundaries of various coupling constants used in astrophysical studies. Here we can do the opposite: we use astrophysical simulations to constrain possible coupling constants and guide nuclear physics experiments.

In this work, we explore, for the first time, the implications of including the \dybnum hexaquark in the description of dense matter within NSs using the CMF model. We analyze its impact on the microphysics of the hybrid EoS and, within the slow hadron-quark conversion regime, we study the possibility of obtaining HSs with the presence of \dyb.
This study is carried out in the framework of modern astrophysical constraints on NSs, which play a key role in our analysis.

The paper is structured in the following manner. In Section \ref{sec:cmf-stab}, we provide the theoretical context of our work. Subsection \ref{subsec:CMFreparametrization} offers a detailed description of the re-parametrization of the CMF model used in this study. In addition, relevant aspects of the stability analysis of HSs are presented in Subsection \ref{subsect:pt}. We focus our attention on the impact of the hadron-quark conversion speed and describe the family of SSHS that appear in the slow conversion regime. In Section \ref{sec:CMF+d*} we discuss the impact of including the \dybnum in the purely hadronic CMF model. We then explore the case where the hadron-quark phase transition is allowed within the CMF model including hexaquarks. Moreover, we present the most relevant astrophysical implications of the appearance of the \dybnum in the cores of HSs and provide an analysis of the constraints imposed by current astronomical observations on the coupling constants associated with this particle.  The most important results regarding the impact of \dyb in HSs constructed within a re-parametrized CMF model are presented in Section \ref{sec:results}. Finally, in Section \ref{sec:summary}, we summarize our main findings and present the most relevant conclusions drawn from our research. Additionally, Appendix~\ref{appendix} describes relevant equations for the CMF model and the particular parametrization that we develop.

\section{Chiral Mean Field model and the slow hadron-quark conversion regime}
\label{sec:cmf-stab}

The CMF model is a theoretical framework developed to study nuclear matter at saturation density and beyond, incorporating essential features of QCD, such as chiral symmetry breaking, broken scale invariance, and a deconfinement transition~\cite{Dexheimer:2010ana}. It is based on a non-linear realization of chiral symmetry~\cite{Weinberg:1968de, Coleman:1969sm, Papazoglou:1997uw} to capture interactions between baryons, mesons, and quarks through the exchange of meson mean fields. This symmetry structure allows the model to account for spontaneous breaking and partial restoration of chiral symmetry under extreme density and/or temperature conditions, which are crucial aspects of QCD in both nuclear and astrophysical contexts. 

In this work, we performed a re-parametrization of quark matter within the CMF model in a zero-temperature approximation, $T=0$, which is particularly relevant for describing matter in fully evolved NSs. This re-parametrization ensures that the EoS containing the entire baryonic octet satisfies the constraints from massive pulsars with masses $M\ge 2\,M_\odot$ before reaching the deconfinement threshold density.

Beyond that point, the dynamical stability and structure of HSs are  determined by the dynamical hadron-quark conversion regime that takes place in the deconfinement region. The hypothesis of \textit{slow} hadron-quark conversion enables the re-parametrized CMF to satisfy astrophysical constraints from NSs that would otherwise be unattainable within the hadronic parametrization we studied.

In the following subsections, we discuss these two aspects in detail.

\subsection{Re-parametrization of the CMF model}
\label{subsec:CMFreparametrization}

Starting from the CMF Lagrangian density within the mean-field approximation
\begin{align}
\mathcal{L}_{\rm CMF}&=\mathcal{L}_{\rm kin}+\mathcal{L}_{\rm 
 int}+\mathcal{L}_{\rm  scal}+\mathcal{L}_{\rm  vec}+ \mathcal{L}_{m_0}+\mathcal{L}_{\rm esb} + \mathcal{L}_{\Phi}\,.
 \label{eq:L_CMF}
\end{align}
we can identify, ${\cal L}_{\rm kin}$, the kinetic term of spin 1/2 fermions (the baryon octet and the three light quarks), ${\cal L}_{\rm int}$, the interaction terms for these fermions with vector ($\omega$, $\rho$, $\phi$) and scalar ($\sigma$, $\delta$, $\zeta$) mesons, and ${\cal L}_{\rm scal}$, the self-interactions among scalar mesons. The term ${\cal L}_{\rm vec}$ contributes to vector meson masses and includes quartic self-interaction terms. The $\mathcal{L}_{m_0}$ term is included to properly fit the compressibility in the C4 parametrization for the vector self-interactions. Additionally, the explicit chiral symmetry breaking term  $\mathcal{L}_{\rm esb}$  allows the model to generate Goldstone boson masses and reproduce the observed hyperon potentials. $\mathcal{L}_{\Phi}$ includes the coupling of Polyakov loop-inspired field $\Phi$ to the fermions, as well as $U_{\Phi}$, the potential term associated to the deconfinement mechanism in the CMF model.

The explicit form of the different Lagrangian terms and detailed descriptions of the model are provided in the Appendix~\ref{appendix}. For a comprehensive derivation of the CMF model, we refer the reader to Ref.~\cite{CruzCamacho:2024psi}.

The in-medium mass for baryons and quarks has an explicit dependence on the scalar field $\Phi$, see \cref{eq:emB,eq:emq}. The scalar field $\Phi$ acts as an order parameter and, therefore, works as a regulating factor to suppress quarks in the hadronic regime and vice versa.

The hadronic and quark total contributions to the energy density, pressure, and baryon number density can be written respectively as
\begin{align}
\varepsilon&=\varepsilon_{\rm fermions}+\varepsilon_{\rm int}+\varepsilon_{\rm bosons}{-P_\Phi}+\frac{\partial P_\Phi}{\partial\mu_B}\mu_B ,\nonumber\\
P&=P_{\rm fermions}+P_{ \Phi}\,,\nonumber\\
n_B&=n_{B,{\rm fermions}}+\frac{\partial P_\Phi}{\partial\mu_B}\,,
\label{eq:net_energy_pressure_density}
\end{align}
where $\mu_B$ is the baryonic chemical potential.  It is noteworthy that in the current version of the model, mesons (both pseudoscalar and vector) are included as non-interacting particles, meaning their masses remain unchanged within the medium (see Ref.~\cite{Kumar:2025rxj} for a different scenario). When charge neutrality and $\beta$ equilibrium with leptons are included to describe the interior of NSs, electrons and muons are added as a free Fermi gas.

\begin{figure}[t!]
    \centering
    \includegraphics[width=0.99\linewidth]{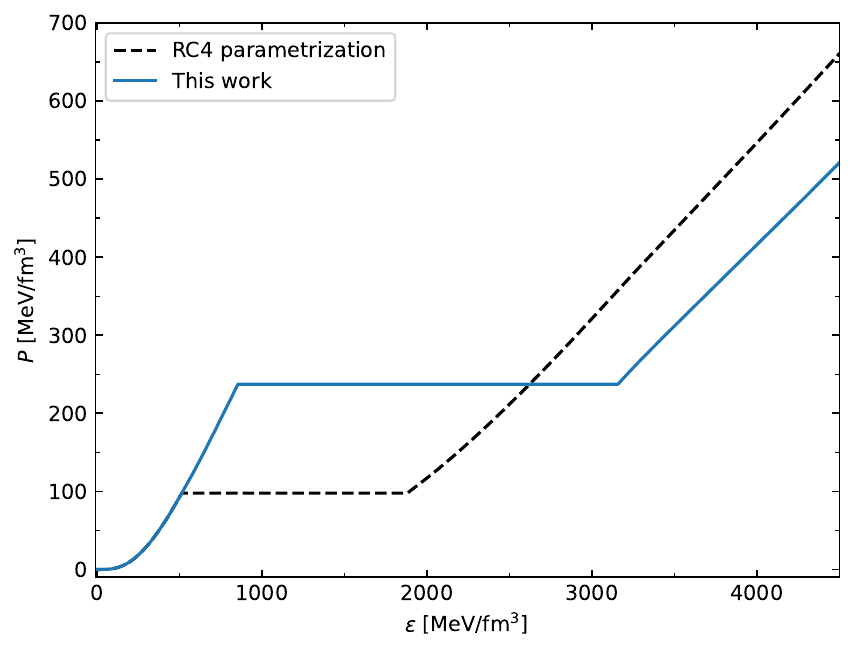}
    \caption{
    Pressure, $P$, as a function of energy density, $\varepsilon$, for the re-parametrized CMF (blue lines)  and RC4 parametrization (black lines). The first-order phase transition induces an energy density gap $\Delta\varepsilon = 2303.58$~MeV/fm$^3$ ($\Delta \varepsilon = 1370.87$~MeV/fm$^3$) and a transition pressure \mbox{$P^{\rm{t}}=237.23$~MeV/fm$^3$} ($P^{\rm{t}}=97.85$~MeV/fm$^3$) for the re-parametrized CMF from this work (original RC4 parametrization).}
    \label{fig:eos-cmf}
\end{figure}

For this work, we make use of the C4 vector meson parametrization within the field redefinition performed in~\citet{Kumar:2024mna} (RC4, where ``R" stands for field redefined) which allows a precise adjustment of the vector meson masses and coupling strengths related to vector meson interactions, while improving the reproduction of key features of the phase diagram as per modern constraints.

\begin{figure}[t!]
    \centering
    \includegraphics[width=0.95\linewidth]{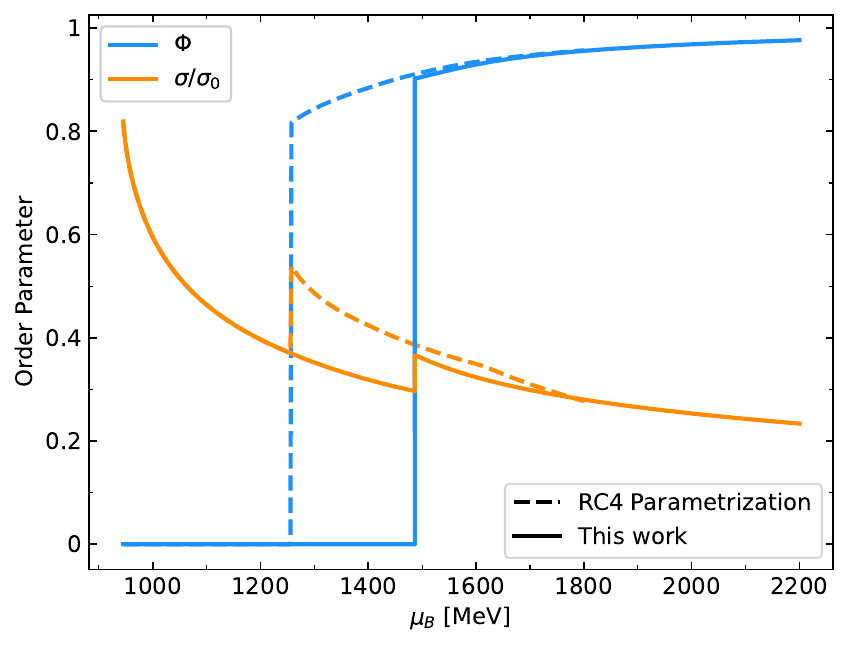}
    \caption{ 
    Order parameters, $\Phi$, related to deconfinement, and $\sigma/\sigma_0$, related to chiral symmetry restoration,  as functions of the baryon chemical potential, $\mu_B$. The first-order deconfinement phase transition is clearly observed at $\mu_B \simeq 1500$ MeV for the re-parametrized CMF model (solid lines) and at $\mu_B \simeq 1250$ MeV for the RC4 parametrization (dashed lines).
}
    \label{fig:phi-cmf}
\end{figure}

We then implemented two modifications in the RC4 CMF parametrization. First, we increased the value of $g_{\Phi q}$ to prevent hadrons from reappearing at high densities. This behavior was not a problem in the original model, as it occurs at extremely high densities, beyond the last stable stellar configuration. However, since our goal is to study the possibility of SSHSs, which can reach extremely high densities in their cores (see details in the following subsection), it was necessary to eliminate this behavior. 
Additionally, the value of the parameter $a_1$ in the potential $U_\Phi$ was increased (decreased in absolute value), to shift the deconfinement phase transition to higher densities. The RC4 parametrization of CMF features a phase transition at $n_b \sim 3\,n_0$,  and a deconfinement phase transition pressure of \mbox{$P^{\rm{t}}_{RC4}=97.85$~MeV/fm$^3$}. This results in a maximum mass of \mbox{$M = 1.871~M_\odot$} (when subjected to an early deconfinement phase transition), which is below of the $2.01\, M_\odot$ required to satisfy the mass value obtained for the pulsar J0740+6620 \citep{Riley:2021anv, Miller:2021tro} . 

Thus, the following two main changes were implemented in this work 
\begin{align}
    g_{\Phi q}^{RC4} &= 470 ~\text{MeV} \xrightarrow{} g_{\Phi q}^{\rm{New}} = 1000 ~\text{MeV} \, ,
\nonumber\\
    a_1^{RC4} &= -1.81 \times 10^{-3} \xrightarrow{} a_1^{\rm{New}} = -1.1 \times 10^{-3} \ .
\end{align}
By adjusting these parameters, we achieved a phase transition at $n_b = 4.6\, n_0$ and a transition pressure of \mbox{$P^{\rm{t}}=237.23$~MeV/fm$^3$.} This phase transition induces an energy density gap of \mbox{$\Delta\varepsilon = $ 2303.58 MeV/fm$^3$} (Fig. \ref{fig:eos-cmf}), which is significantly larger than the energy density gap 
 \mbox{$\Delta \varepsilon_{RC4} = 1370.87$~MeV/fm$^3$ }obtained using the RC4 parametrization.
 
Following these modifications, the order parameters $\sigma/\sigma_0$ and $\Phi$ do not exhibit qualitative changes, both maintaining the behavior observed in the RC4 parametrization.  
 As shown in Fig.~\ref{fig:phi-cmf}, the chiral condensate, $\sigma$, decreases from its vacuum value, indicating chiral symmetry restoration, except for the baryon chemical potential where $\Phi$ jumps from $\sim 0$ to $\sim 1$ (corresponding to the first-order deconfinement phase transition), where a small first-order phase transition is also induced. In the context of the new parametrization, both discontinuities occur at higher baryon chemical potentials, specifically at $\mu_B \simeq 1500$ MeV, due to the delayed first-order phase transition.

\begin{figure}[t!]
    \centering
    \includegraphics[trim={0 0 0 1cm},clip,width=1.04\linewidth]{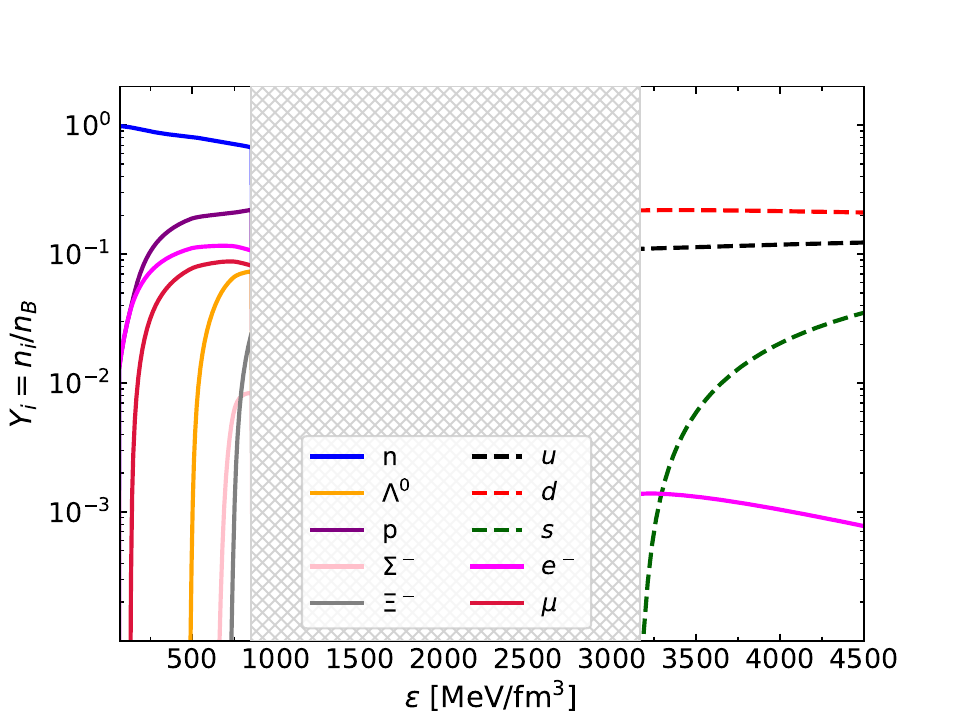}
    \caption{
    Particle fractions, $Y_i = n_i / n_B$, as a function of the energy density, $\varepsilon$, for the re-parametrized CMF model. Here, $i$ refers to the particle species, and $n_B$ is the baryon number density. With the new parameters, hadrons do not reappear at extremely high energy densities. The gray area represents the energy density gap of $\Delta\varepsilon = $ 2303.58 MeV/fm$^3$ shown in the hybrid EoS of Fig.~\ref{fig:eos-cmf}.}
    \label{fig:pop-cmf}
\end{figure}

The particle fractions, presented in Fig.~\ref{fig:pop-cmf}, show the most noticeable changes at extremely high energy densities (and baryon chemical potential). With the increase in $g_{\Phi q}$, hadrons do not reappear at least until \mbox{$\varepsilon = 4500$ MeV/fm$^3$}. This new behavior allows for the formation of extremely compact HSs without hadrons reappearing in the quark phase.

Regarding the mass-radius relationship, depicted in Fig.~\ref{fig:mr-cmf}, the most significant impact is observed in the maximum mass achievable before the kink that occurs when stars cross the threshold density of \mbox{$n_b = 4.6\, n_0$}, where deconfinement takes place. The red circles on each curve mark the threshold for quark deconfinement. The maximum mass configuration increases from \mbox{$M = 1.871 M_\odot$} to \mbox{$M = 2.118 M_\odot$} (when attaching a low-density BPS crust \cite{Baym:1971pw} to each EoS before solving the TOV equations). 
As shown in this figure, the mass-radius relationship  for this work now satisfies the mass constraint imposed by the observation of the pulsar J0740+6620, \mbox{$M_{\rm{max}} \geq 2.01 M_\odot$.} The stability of stellar configurations beyond the maximum mass, for which $\partial M/\partial \epsilon _c < 0$, is discussed in the following subsection. 

\subsection{Phase transition and slow stability branch}
\label{subsect:pt}

As already described in the introduction, the nature of the hadron-quark phase transition is mainly governed by the (currently unknown) value of the surface tension at the hadron-quark interface. It is generally agreed that if the surface tension exceeds a critical value, the Maxwell construction for the phase transition (in which a discontinuity in the energy density occurs at a fixed pressure) is favored \cite{Voskresensky:2002csa,  Endo:2011roh,Wu:2018eoq}, resulting in a sharp phase transition. Conversely, if the surface tension is below this critical value, a mixture of phases is energetically favored. In the limit of very low surface tension,
a Gibbs construction for the phase transition, as proposed by Glendenning in Ref. \cite{Glendenning:1992fop}, can be considered (see, for example, Refs. \cite{Endo:2006cse,hempel:2013not,Wu:2019nse} and references therein).  

For HSs with a sharp phase transition, the nucleation timescale, that is, the speed at which hadrons dissolve into quarks (and vice versa),  plays a fundamental role in their stability against radial oscillations, i.e. in its dynamical stability \cite{Pereira:2018pte,mariani:2019mhs,Malfatti:2020dba,Tonetto:2020dgm,rodriguez:2021hsw,Goncalves:2022ios,Mariani:2022omh,Ranea:2022bou,lugones:2023ama,Ranea:2023auq,Ranea:2023cmr,Rau:2023tfo,Rau:2023neo, Rather:2024roo}. 
Although the nucleation timescale remains unknown, its impact on HSs has been studied considering two regimes: \textit{slow} and \textit{rapid}. If the nucleation timescale is much longer than the characteristic timescale of radial perturbations, we refer to it as \textit{slow}. The opposite case represents the \textit{rapid} regime.

\begin{figure}[t!]
    \centering
    \includegraphics[width=0.98\linewidth]{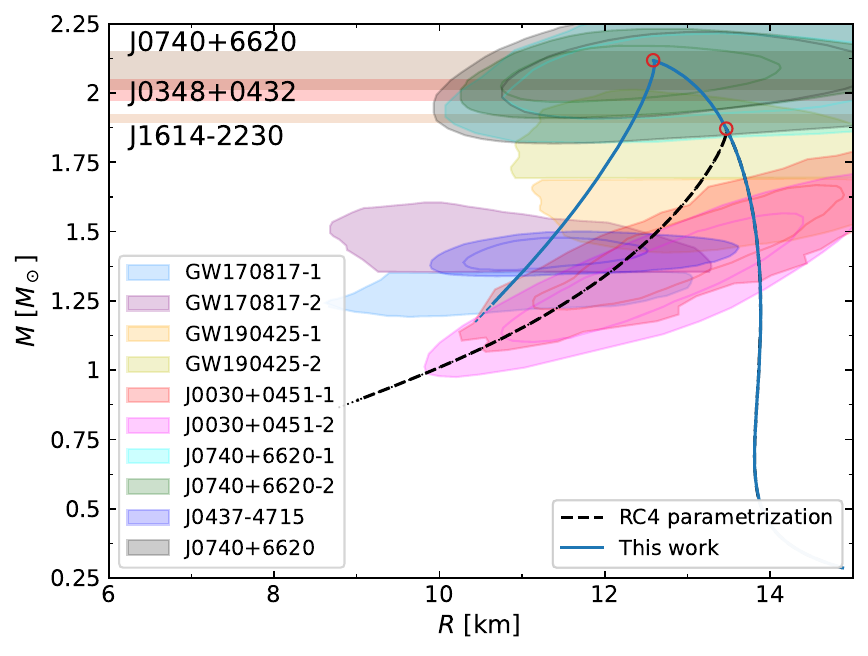}
    \caption{
    Mass-radius diagram for the hybrid EoSs shown in Fig.~\ref{fig:eos-cmf}. Stellar configurations constructed under the slow conversion hypothesis at the hadron-quark interface are represented for the re-parametrized CMF model and for the RC4 parametrization. Red circles on each curve mark the threshold for quark deconfinement. The shaded areas represent recent astrophysical observations. It can be seen that the re-parametrized CMF model mostly satisfies all modern astrophysical constraints for NSs displayed in the diagram  when the extended stability branch is considered.
}
    \label{fig:mr-cmf}
\end{figure}

Calculations of radial oscillations without considering nucleation times lead to the well-known necessary condition for stability, \mbox{$\partial M/\partial \varepsilon _c > 0$}, where the limit corresponds to the maximum mass configuration of the star. This result arises from the fact that for configurations in the \mbox{$\partial M/\partial \varepsilon _c > 0$} branch  every radial eigenfrequency is a real value. However, further analysis incorporating nucleation times has shown that, in the case of \textit{slow} phase transitions, stellar configurations can remain stable against linearized radial perturbations beyond the maximum mass \citep{Pereira:2018pte}. These stars, known as SSHS \cite{lugones:2023ama, Mariani:2024cas}, form an extended stability branch. 

Within the context of our work, considering cold and catalyzed matter, it is important to note that the existence of this extended stability branch is only possible in the presence of a \textit{slow} sharp first-order phase transition. In cases with a purely hadronic EoS (or no deconfinement before the maximum mass configuration), SSHSs cannot form.
In particular, Fig.~\ref{fig:mr-cmf} shows the SSHS branch extending beyond the maximum mass configuration toward smaller radii, up to the end of the continuous curve; this scenario allows for the existence of a large family of twin HSs that would otherwise not exist. The blue curve constructed with the re-parametrized CMF model, when considering the extended stability branch, mostly satisfies all modern astrophysical constraints for NSs displayed in the mass-radius diagram. In the following analysis, we will adopt the slow conversion hypothesis and, consequently, explore the implications of SSHSs when considering hybrid EoSs.

\section{\dybnum hexaquarks in the re-parametrized CMF} \label{sec:CMF+d*}

Recently, the inclusion of hexaquarks has been explored to study dense matter within various parametrizations of the density-dependent RMF model \cite{Celi:2024doh}. In that model, the hexaquark mass is governed by the $\sigma$ meson (which mediates attractive nucleon-nucleon interactions), while its baryon chemical potential depends on the $\omega$ meson (which generates repulsive forces between nucleons at short distances). 

In this work, following the approach described in Refs.~\cite{Mantziris:2020nsm, Celi:2024doh}, we adopt a simplified treatment of hexaquarks. The Lagrangian for the hexaquarks is given by
\begin{equation}
\mathcal{L}_{d^*}=\mathcal{D}^* \xi_{d^*}^* \mathcal{D} \xi_{d^*}-m_{d^*}^{* 2} \xi_{d^*}^* \xi_{d^*} \, ,
\label{eq:Ld*}
\end{equation}
where $\mathcal{D}=\left(\partial_\mu+i g_{\omega d^*} \omega_\mu\right)$, $m_{d^*}^*=m_{d^*}-g_{\sigma d^*} \sigma$ corresponds to the effective mass and $\xi_{d^*}$  is the scalar-isoscalar field of the hexaquark. Following the prescription of Ref.~\cite{Celi:2024doh}, we treat a spin $S=3$ \dyb state as a $S=0$ particle. Note that, to align with experimental results, we need to set $m_{d^*}^* = 2380$ MeV in vacuum. This requires adjusting the value of $m_{0}^{d^*}$. Since $m_{d^*}^*$ depends on the coupling constants, its vacuum value changes depending on the chosen coupling values. Thus, \mbox{$m_{0}^{d^*}= 2380 $ MeV$ - g_{\sigma d^*}\sigma_0$}.

By adding the \dyb contribution, \Cref{eq:Ld*}, to the CMF Lagrangian, \Cref{eq:L_CMF}, we obtain the modified set of coupled equations of motion given by
\begin{widetext}
\begin{align}
\label{eq:sigma}
\sigma: \quad 0 & =-\sum_{i \in {\rm fermions}} g_{ \sigma i} {n_{S i}} + g_{\sigma d^*} n_{d^*}- k_0 \chi^2 \sigma+4 k_1\left(\sigma^2+\delta^2+\zeta^2\right) \sigma+2 k_2\left(\sigma^2+3 \delta^2\right) \sigma  
+2 k_3 \chi \sigma \zeta \nonumber \\
& \qquad  +\frac{2 \varepsilon}{3} \chi^4 \frac{\sigma}{\sigma^2-\delta^2}-m_\pi^2 f_\pi\left(\frac{\chi}{\chi_0}\right)^2 \,,   \\
\label{eq:delta}
\delta: \quad 0 & =-\sum_{i \in {\rm fermions}} g_{ \delta i} {n_{S i}}   -  k_0 \chi^2 \delta+4 k_1\left(\sigma^2+\delta^2+\zeta^2\right) \delta+2 k_2\left(3 \sigma^2+\delta^2\right) \delta  -2 k_3 \chi \delta \zeta-\frac{2 \varepsilon}{3} \chi^4 \frac{\delta}{\sigma^2-\delta^2}\,, \\
\label{eq:zeta}
\zeta: \quad 0 & =-\sum_{i \in {\rm fermions}} g_{ \zeta i} {n_{S i}} - k_0 \chi^2 \zeta+4 k_1\left(\sigma^2+\delta^2+\zeta^2\right) \zeta+4 k_2 \zeta^3+k_3 \chi\left(\sigma^2 -\delta^2\right) \nonumber \\
& +\frac{\varepsilon}{3 \zeta} \chi^4 - \left(\sqrt{2} m_k^2 f_k-\frac{1}{\sqrt{2}} m_\pi^2 f_\pi\right)\left(\frac{\chi}{\chi_0}\right)^2 \,, \\
\label{eq:omega}
 \omega: \quad 0 & =-\sum_{i \in{\rm fermions}} g_{ \omega i} {n_i} +  g_{\omega d^*} n_{d^*} + \left(\frac{\chi}{\chi_0}\right)^2 m_\omega^2 \omega + g_4 \left(4 \omega^3 + 6 \phi^2 \omega+6 \sqrt{2} \phi \omega^2+\sqrt{2} \phi^3\right) \,,
 \\
\label{eq:rho}
\rho: \quad 0 & =-\sum_{i \in {\rm fermions}} g_{ \rho i} {n_i} + \left(\frac{\chi}{\chi_0}\right)^2 m_\rho^2 \rho\,, \\
\label{eq:phi}
\phi: \quad 0 & =-\sum_{i \in {\rm fermions}} g_{ \phi i} {n_i} + \left(\frac{\chi}{\chi_0}\right) m_\phi^2 \phi+g_4\left(2\sqrt{2}\sqrt{\frac{Z_\phi}{Z_\omega}}\omega^3 + 6\left(\frac{Z_\phi}{Z_\omega}\right)\omega^2\phi + 3\sqrt{2}\omega\phi^2\left( \frac{Z_\phi}{Z_\omega} \right)^{3/2} + \left( \frac{Z_\phi}{Z_\omega} \right)^{2}\phi^3\right)\,,\\
\label{eq:poly}
\Phi: \quad 0 & =-\sum_{i \in \rm baryons} 2 g_{\Phi B} {n_{S i}}   \Phi-2 g_{ \Phi d^*} {n_{d^*}   \Phi}+\sum_{i \in \rm quarks}  g_{ \Phi q}       {n_{S i}}     +2\left(a_0 T^4+a_1 \mu_B^4+a_2 T^2 \mu_B^2\right) \Phi  +a_3 T_0^4 \frac{12 \Phi}{3 \Phi^2-2 \Phi-1}\,,
\end{align}
\end{widetext}
where only the equations for $\sigma$ and $\omega$ are modified, and
\begin{equation}
    n_{d^*}=2 (m_{d^*} - g_{\sigma d^*} \bar{\sigma}) \xi^*_{d^*}\xi_{d^*}=2 (\mu_{d^*} + g_{\omega d^*} \bar{\omega}) \xi^*_{d^*}\xi_{d^*}\,,
\end{equation}
is the number density of the \dyb hexaquark, being \mbox{$\mu_{d^*} = 2 \mu_n - \mu_e$} \cite{Mantziris:2020nsm}. Note that in this work, we assume the frozen glueball limit, $\chi$=$\chi_0$~\cite{CruzCamacho:2024psi}.

\begin{figure}[t!]
    \centering
    \includegraphics[width=0.9\linewidth]{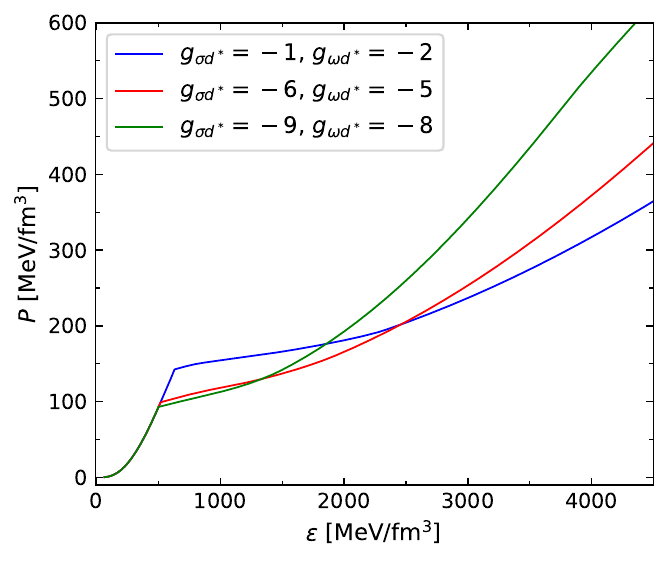}
    \caption{
    Pressure, $P$, as a function of the energy density, $\varepsilon$, for three different sets of coupling constants, chosen to represent different regions of the parameter space, including the \dyb  with couplings in the range given by Eq.~(\ref{coupling_d*}). From now on, we only show results for the reparametrized CMF from this work. It can be seen that the appearance of the \dyb flattens the EoS, causing a sudden drop in pressure. This result is in agreement with the previous works by \citet{Celi:2024doh, Mantziris:2020nsm}.}
    \label{fig:eos-hadronic}
\end{figure}

\begin{figure*}[t!]
    \centering
    \includegraphics[width=0.95\linewidth]{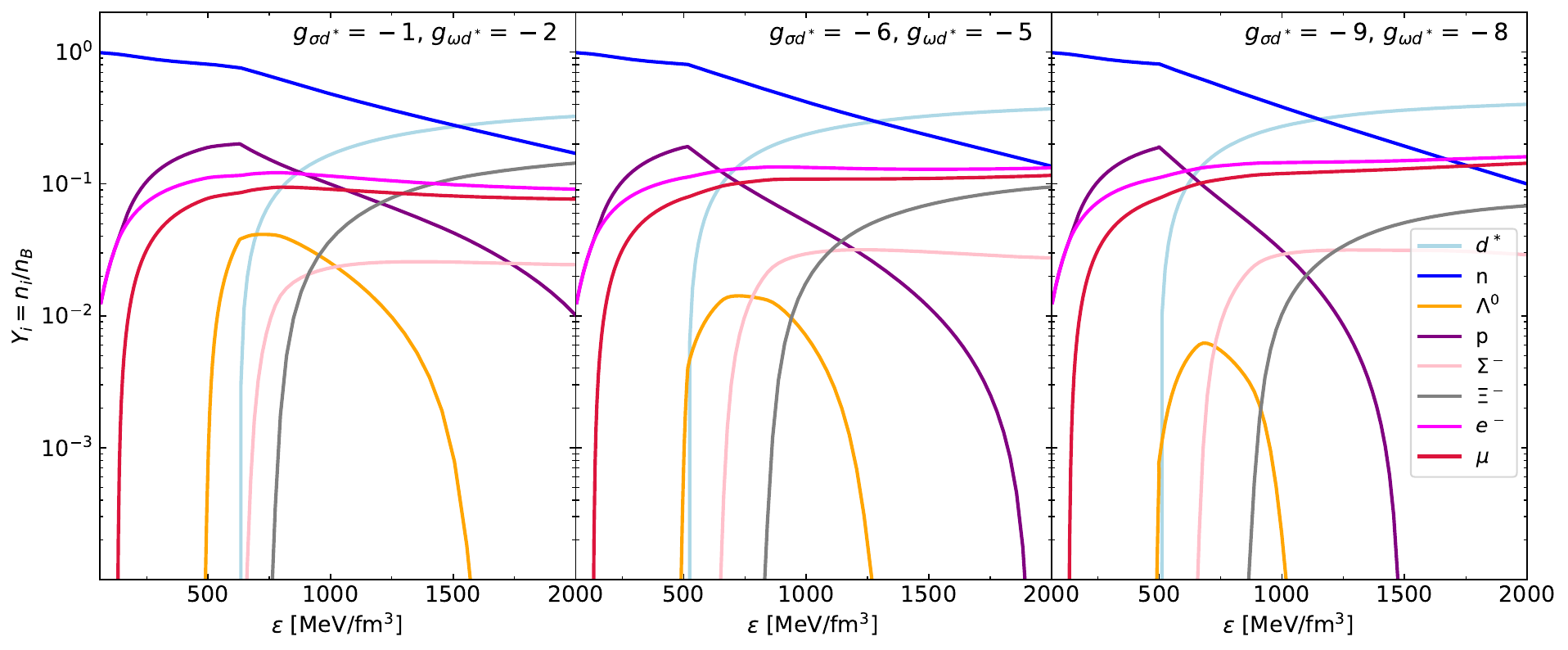}
    \caption{ 
    Particle fractions, $Y_i$, as a function of energy density, $\varepsilon$, for the three EoSs shown in Fig.~\ref{fig:eos-hadronic}. The three sets of coupling constants chosen lead to qualitatively similar scenarios. In the range \mbox{$500~\rm{MeV/fm}^3 \lesssim \varepsilon \lesssim 625\, \rm{MeV/fm}^3$}, \dyb hexaquarks appear, rapidly increasing their abundance and quickly becoming the most abundant particles. This result is also in agreement with previous studies by \citet{Celi:2024doh, Mantziris:2020nsm}. It can be seen that as the coupling constants increase in magnitude, the presence of protons and $\Lambda^0$ decreases significantly for high values of energy density.}
    \label{fig:pop-had}
\end{figure*}

\begin{figure}[t!]
    \centering
    \includegraphics[width=0.95\linewidth]{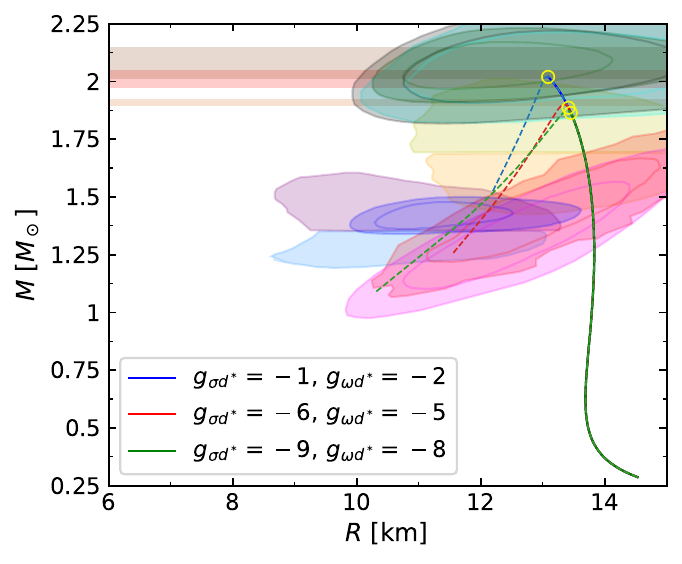}
    \caption{ 
    Mass-radius diagram for the three EoSs shown in Fig.~\ref{fig:eos-hadronic}. Solid lines represent stable configurations, while dashed lines represent unstable ones (the circles mark the threshold of \dyb appearance). An earlier emergence of hexaquark \dyb leads to an earlier destabilization (lower density threshold) of stellar configurations and, consequently, lower stellar maximum masses. This is also in agreement with the studies by \citet{Celi:2024doh, Mantziris:2020nsm}. 
}
    \label{fig:enter-label}
\end{figure}

Let us first evaluate how \dyb impacts the CMF in the absence of quark degrees of freedom ($\Phi=0$) to fix the intervals of the coupling constants related to \dyb. These constants will then be used to study the \dyb impact when a phase transition to quark matter occurs. Since the RMF model and the CMF model at the hadronic level share fundamental characteristics, such as describing nuclear interactions mediated by scalar and vector mesons (like $\sigma$ and $\omega$), and because the form of the CMF Lagrangian is analogous to that of the RMF, it is expected that the hadronic CMF, when incorporating the hexaquark, exhibits a similar behavior to that observed in the RMF model. This assumption facilitates the determination of an appropriate interval for the coupling constants \gsd and \gwd in the CMF model. From \citet{Celi:2024doh}, the coupling constant ratios $x_{\sigma d^*}$ and $x_{\omega d^*}$ are given by
\begin{equation}
    x_{\sigma d^*} = \frac{g_{\sigma d^*}}{g_{\sigma N}}, \quad x_{\omega d^*} = \frac{g_{\omega d^*}}{g_{\omega N}}\,.
\end{equation}

In \citet{Celi:2024doh}, although the range \(0 \le x_{\sigma d^*} \le 2\) and \(-2 \le x_{\omega d^*} \le 0\) was explored, relevant results were only obtained for \(0 \le x_{\sigma d^*} \lesssim 1\) and \(-1 \lesssim x_{\omega d^*} \le 0\). Considering that in the CMF model \(g_{\sigma N} = -9.83\) and \(g_{\omega N} = 11.9\), the corresponding intervals for \dyb within the CMF model are given by
\begin{equation}  
    -9.83 \le g_{\sigma d^*} \le 0 \quad \text{and} \quad -11.9 \le g_{\omega d^*} \le 0\,.
    \label{coupling_d*}  
\end{equation}
In what follows, we explore this range.

In Fig.~\ref{fig:eos-hadronic} we show the pressure as a function of energy density for the hadronic CMF in the presence of \dyb. To explore the behavior of the CMF model, we investigate the parameter space for the ratio of the coupling constants of the $\sigma$ and $\omega$ mesons to the \dyb hexaquark, with values ranging as given in Eq.~(\ref{coupling_d*}). Each curve represents a specific combination of \gsd and \gwd. The behavior of the EoSs is similar to that reported in \citet{Celi:2024doh}: since the \dyb hexaquark is a boson, its appearance leads to an abrupt \textit{softening} of the EoS, i.e., the increase in pressure is reduced with increasing energy density.

Furthermore, in Fig.~\ref{fig:pop-had} we present the particle fractions as a function of the energy density for the EoSs shown in Fig.~\ref{fig:eos-hadronic}. It can be seen that, depending on the coupling constants associated with the hexaquarks, such particles (light blue lines) appear at energy densities in the range \mbox{$500~\rm{MeV/fm}^3 \lesssim \varepsilon \lesssim 625\, \rm{MeV/fm}^3$}. When the hexaquarks appear, the fractions of $e^-$ and $\mu$ increase  slightly to maintain charge neutrality, as the \dyb is positively charged. Moreover, since it has a baryon number of $2$, proton and neutron fractions (blue and maroon lines, respectively) decrease significantly upon its appearance. Furthermore, it can be seen that as the coupling constants increase in magnitude, the presence of protons and $\Lambda^0$ decreases noticeably at high energy densities. This feature aligns with the EoSs behavior in Fig.~\ref{fig:eos-hadronic}, showing stiffer EoSs for high densities as the coupling constants increase. Hence, the reason for this stiffening appears to be the reduction of particle species degrees of freedom through the decreasing and disappearance in these particle fractions.

In Fig.~\ref{fig:enter-label} we show the mass-radius relationship corresponding to the EoSs presented in Fig.~\ref{fig:eos-hadronic}. The bosonic nature of the \dyb particle and the reduction in the nucleon population due to its appearance coincide with the flattening of the EoSs and, consequently, with a dramatic reduction in the maximum stellar mass in the mass-radius plane. Although the green line EoS of Fig.~\ref{fig:eos-hadronic} stiffens at high densities, this effect is not enough to achieve a higher maximum mass in the mass-radius diagram, since \dyb appears earlier when the coupling constants increase in magnitude, as shown in Fig.~\ref{fig:pop-had}. The \dyb presence destabilizes NSs, in agreement with previous studies \cite{Mantziris:2020nsm,Celi:2024doh}.

We would like to clarify that for the present analysis, we have chosen not to include $\Delta$ resonances to maintain a clear focus on the impact of the \dybnum hexaquark on the hadron-quark phase transition ($\Delta$ resonances plus \dyb have been already included and analyzed in purely hadronic stars, in Ref.~\cite{Celi:2024doh}). Additionally, the inclusion of $\Delta$ baryons would significantly expand the parameter space to be analyzed, making the analysis too complex and potentially obscuring one of the main focuses of this work: the role played by \dybnum in the deconfinement phase transition.

\section{Hybrid stars containing \dybnum}
\label{sec:results}

Once results consistent with previous works involving \dybnum have been obtained, we look for the possibility of a first-order phase transition. To achieve this, we combine the CMF re-parametrization of Subsection~\ref{subsec:CMFreparametrization} with the treatment for hexaquark inclusion presented in Section~\ref{sec:CMF+d*}. Using this hybrid EoS model, we systematically and extensively explore its parameter space consisting of only two free parameters: the coupling constants of the \dyb particle \gwd and \gsd. Taking into account Eq.~\eqref{coupling_d*}, we construct a fine grid with a step size $\Delta g_{i \dyb}=0.25$, within the following range:
\begin{equation}
    -10 \le g_{\sigma \dyb} \le 0 \,,\;   -10 \le g_{\omega \dyb} \le 0 \,.
    \label{coupling_d*_range}
\end{equation}
For each of these sets, we construct the EoS, obtaining the pressure-energy density relationship and the particle populations. We then attach a low-density BPS-BBP crust \cite{Baym:1971pw} to each EoS and solve the TOV equations under the slow hadron-quark conversion speed scenario to analyze the stability of these stellar configurations against radial perturbations. This latter hypothesis, in the case of hybrid EoSs, gives rise to the SSHS stability branch, which extends beyond the maximum mass configuration towards smaller radii, up to the terminal mass configuration where unstable configurations appear.

\begin{figure*}
    \centering
\includegraphics[width=0.95\linewidth]{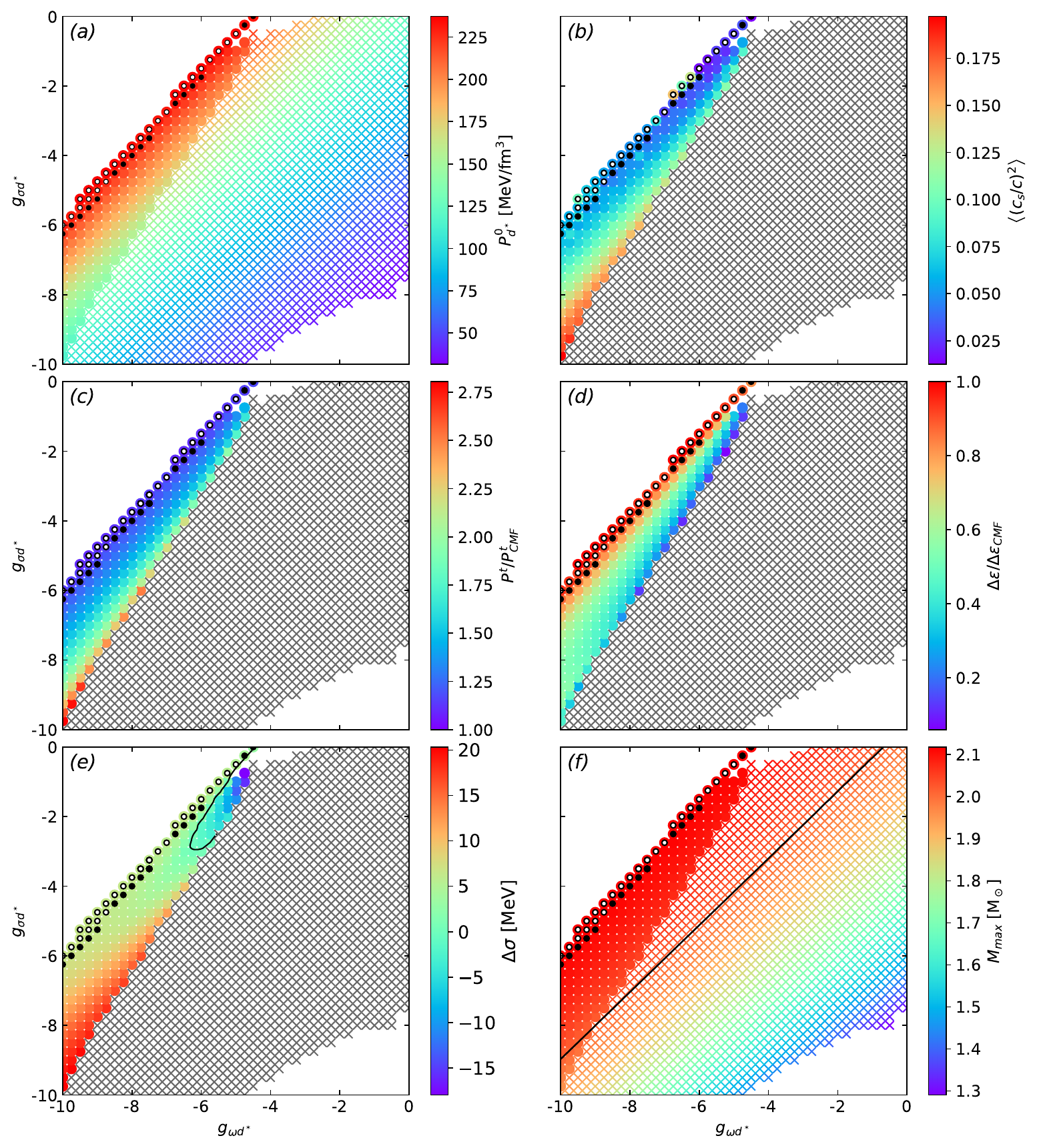}
    \caption{Three-dimensional plots showing the \dyb coupling constants (\gwd, \gsd) on two axes and relevant physical quantities indicated by color code. Panel (a): Pressure of \dyb appearance. Panel (b): Mean value of squared speed of sound in the region bounded by $P^0_{d^*}$ and $P^{\rm{t}}$. Panel (c) Deconfinement pressure normalized by the deconfinement pressure for the model without \dyb. Panel (d): Deconfinement energy density jump normalized by energy density jump for the model without \dyb. Panel (e): Gap in the $\sigma$ value after the phase transition (the black curve corresponds to  $\Delta \sigma = 0$). Panel (f): Maximum mass (the black line marks the $2.01 \,M_\odot$ limit).   
    Crosses forming a \textit{grid} represent physically acceptable solutions without a phase transition. Black crosses indicate regions where the color-mapped quantity is undefined. Colored filled dots represent solutions that are both physically acceptable and compatible with a phase transition. Black dots mark configurations that allow SSHSs with \dyb presence, and those with a white dot inside are solutions compatible with standard astrophysical observations of NSs.
}
    \label{fig:omega-sigma}
\end{figure*}

Based on these results, we investigate which sets result in purely hadronic EoSs (without a  phase transition) and which sets yield hybrid hadron-quark EoSs. In addition, we determine in which of the hybrid sets a \dyb population  appears in the hadronic phase. In Fig.~\ref{fig:omega-sigma}, we present these results in the \gwd-\gsd plane. In the different three-dimensional panels, the crosses indicate the sets that correspond to purely hadronic EoSs (no phase transition), \textit{i.e.} the solutions discussed in Subsection~\ref{sec:CMF+d*}. Colored circles indicate the sets where a phase transition occurs after the \dyb appearance, \textit{i.e.} hybrid EoSs that include hexaquarks in the hadronic phase. 
Blank spaces represent regions where no solutions can be found: two large triangular regions in the upper left and lower right corners, and a small irregular blank region in the upper middle section. 
The blank upper triangle corresponds to the region where the hadron-quark phase transition occurs before the hexaquark appearance, \textit{i.e.} \pdyb$>P^{\rm{t}}$, being \pdyb the pressure at which hexaquarks appear and $P^{\rm{t}}$ the hadron-quark phase transition pressure. Although these solutions are physically valid, they are irrelevant to our study as they do not include \dyb population, and  we therefore discard them. The sets in the blank lower triangle are also discarded, as they represent non-physical solutions due to anomalous behaviors of the \dyb particle, such as non-monotonic energy density and/or pressure. 
The small blank region in the upper middle corresponds to sets with numerical instabilities that prevent obtaining reliable results.
 
The large central region of the panels in Fig.~\ref{fig:omega-sigma} is filled with a crossed grid, marking the EoSs that include hexaquarks but no quark phase. Most of the studied sets belong to this family. Finally, a narrow triangular region defined by the colored circles represents the parameter space where EoSs include both hexaquark and a quark phase. The black and white small circles correspond to subsets with astrophysical implications, which are explained and discussed later. Each panel also includes a color map for crossed and circle sets, representing different relevant quantities. Some quantities are undefined in the absence of deconfinement; therefore, for these panels, the crossed sets appear in black.

In panel (a) of Fig.~\ref{fig:omega-sigma}, the color map indicates the pressure at which hexaquarks appear, \pdyb; it can be seen that a wide range of \pdyb values can be achieved by varying the coupling constants, with higher \pdyb values obtained when \gwd decreases and when \gsd increases. Also, it is evident that for a first-order hadron-quark phase transition to occur, the hexaquark must not appear too early, requiring $\pdyb \gtrsim 150$~MeV/fm$^3$. Otherwise, no crossing of the phases is observed in the $P$-$\mu_B$ plane.

In panel (b) of Fig.~\ref{fig:omega-sigma}, we define a new quantity: the mean value of the squared speed of sound in the region where \dyb is present, $\pdyb < P < P^{\rm{t}}$, expressed in units of the speed of light \cs. This quantity allows one to characterize the morphology of the \dyb contribution to the EoS, indicating the mean slope of this segment of the $P$-$\varepsilon$ relationship. As the sets marked with crosses do not reproduce a phase transition, \cs (mean value of squared speed of sound in the region bounded by $P^0_{d^*}$ and $P^{\rm{t}}$) is undefined for these cases, and they are shown in black. In general, the presence of hexaquarks induces a dramatic \textit{softening} on the EoS (as indicated by the larger jump in $\varepsilon$ in panel (d)), resulting necessarily in lower values of \cs compared to the traditional speed of sound in hadronic EoSs. Although minor anomalies with discontinuous behavior in the averaged speed of sound are observed, a global trend is evident: \cs decreases when \gwd decreases and when \gsd increases.

In panel (c) of Fig.~\ref{fig:omega-sigma}, we present the same sets, with the color map  now representing the deconfinement phase transition pressure normalized by the corresponding value in the re-parametrized model without \dyb, $P^{\rm{t}}/P^{\rm{t}}_{\rm{CMF}}$, being $P^{\rm{t}}_{\rm{CMF}}=237.23$~MeV/fm$^3$. Sets marked with crosses, which lack a phase transition, are shown in black. Since we only show values for which $P^{\rm{t}}/P^{\rm{t}}_{\rm{CMF}}>1$, an empty region appears in the upper middle section where $P^{\rm{t}}/P^{\rm{t}}_{\rm{CMF}}$ would be less than $1$. 
The presence of hexaquarks not only affects the possibility of the phase transition occurring, but also increases the transition pressure, delaying the appearance of the quark phase compared to the model without \dyb.
In contrast, panel (d) of Fig.~\ref{fig:omega-sigma} shows the energy density gap across the phase transition, normalized to the corresponding value in the model without \dyb, $\Delta \varepsilon /\Delta \varepsilon_{\rm{CMF}}$ (being $\Delta\varepsilon_{\rm{CMF}} = 2303.58 $~MeV/fm$^3$). Here, the presence of \dyb always reduces this gap, with $\Delta \varepsilon /\Delta \varepsilon_{\rm{CMF}}<1$. This reduction in $\Delta \varepsilon$ has interesting astrophysical implications that we analyze below.

In panel (e) of Fig.~\ref{fig:omega-sigma}, the color map represents the discontinuity in the $\sigma$-meson field across the deconfinement phase transition, $\Delta \sigma$. In the CMF model without \dyb, the value of the discontinuity, as shown in Fig.~\ref{fig:phi-cmf}, is \mbox{$\Delta \sigma = 6.5$~MeV}. Sets marked with crosses, which do not exhibit a deconfinement phase transition, are shown in black. In our extensive exploration of the \dyb coupling constants, we find that most sets yield  $\Delta \sigma > 0$, consistent with the model without \dyb. However, we also identify a small region of parameter space where $\Delta \sigma < 0$; in particular, we calculate and display (with a black curve in the figure) the level curve \mbox{$\Delta \sigma = 0$}, which corresponds to sets where the $\sigma$-meson field remains continuous across the hadron-quark phase transition. This result is particularly interesting, as it could indicate not only that the presence of hexaquarks influences the dynamics of the hadron-quark phase transition, but also that hexaquarks may affect the chiral symmetry restoration during the phase transition.

In addition to all the microphysics results presented, we also show the results obtained after solving the TOV equations. In panel (f) of Fig.~\ref{fig:omega-sigma}, we present the maximum mass star for each EoS set, $M_{\rm{max}}$. Given that the re-parametrized CMF model without \dyb produces a maximum mass of \mbox{$M_{max,\, CMF} = 2.118 \, M_\odot$} (see Subsection~\ref{subsec:CMFreparametrization}), it can be noticed that the \dyb particles always reduce this value. This behavior is expected because, as we have already shown, the hexaquark appearance strongly \textit{softens} the EoS. Nevertheless, numerous sets satisfy the maximum mass constraint, $M_{\rm{max}} \geq  2.01 \, M_\odot$. We highlight the level curve $M_{\rm{max}} = 2.01 \, M_\odot$ with a black line, with higher values of $M_{\rm{max}}$ corresponding to larger values of \gwd and smaller values of \gsd. Note that these results are independent of the assumption of the hadron-quark conversion speed or the extended stability branch of HSs.

To thoroughly examine the astrophysical implications of our resulting EoSs, we implement a series of filtering techniques. We first identify which parameter combinations produce stable HSs with \dyb presence, and then determine which of these combinations simultaneously satisfy  all current astrophysical constraints for NSs (discussed in Section~\ref{sec:intro} and presented in Fig.~\ref{fig:mr-cmf}). In Fig.~\ref{fig:omega-sigma}, we present these two subsets: sets that produce stable HSs with the  presence of hexaquarks are highlighted with black dots, while those that also satisfy all astrophysical constraints are marked with an inner white dot. As can be seen in any of the panels in this figure, only a small range of parameter combinations remains after applying the astrophysical filters. Although those sets correspond to hybrid EoSs with \dyb particles, the absence of black dots in most of the colored circles indicates that quark matter appears after NS destabilization,
preventing the existence of stable HSs regardless of the conversion speed analyzed.
Also, it is important to note that all of the stable HSs obtained are SSHSs, \textit{i.e.},  objects belonging to the \textit{slow} extended stability branch. Without the \textit{slow} hadron-quark conversion hypothesis, no stable NSs with a quark core would be obtained, as no HSs exist before the maximum mass configuration. Moreover, it is interesting to note that no white dots appear outside the black dot subset. There should also be noted that this result has an inevitable model dependence, since the purely hadronic branch produced by our EoS parametrization is \textit{stiff} enough not to satisfy the GW170817 and J0437-4715 radii. If other CMF hadronic parametrizations or softer hadron EoSs in the low-density regime were considered, as explored in Ref.~\cite{Celi:2024doh}, the parameter ranges satisfying astrophysical constraints might expand, even without requiring the SSHS branch. In other words, within our model, the existence of SSHSs is necessary to meet the small radius constraints; this result, which highlights the importance of the SSHS branch when considering \textit{stiff} hadron EoS in the context of modern astrophysical constraints, is studied and generalized in Refs.~\cite{lugones:2023ama, Mariani:2024cas}.

As previously discussed and also found in \citet{Celi:2024doh}, the inclusion of the \dyb particle in the EoS always reduces the maximum mass of NSs compared to the hadronic EoS without \dyb and, also, the appearance of the \dyb induces a kink in the $M$-$R$ relationship. Hence, looking simultaneously at panels (a) and (f) of Fig. \ref{fig:omega-sigma}, we can conclude that to achieve a maximum mass of at least $2.01~M_\odot$, \pdyb must not be too low. Furthermore, in view of the objectives of this work considering the possibility of a hadron-quark phase transition, stable hybrid hadron-quark branches require a phase transition pressure, $P^{\rm{t}}$ (presented in panel (c) of Fig.~\ref{fig:omega-sigma}), to be sufficiently close to the \dyb appearance pressure, \pdyb. If $P^{\rm{t}}$ is too far from \pdyb, quark matter appears after the \dyb particles have already destabilized the NS. On the other hand, as shown in panel (d) of Fig.~\ref{fig:omega-sigma}, the inclusion of hexaquarks in the EoS always reduces the energy density gap of the phase transition, $\Delta \varepsilon$. As discussed in Refs.~\cite{lugones:2023ama, Mariani:2024cas}, longer SSHS branches are produced for larger  values of $\Delta \varepsilon$. Since our model needs long enough extended branches to satisfy the GW170817 and J0437-4715 constraints, $\Delta\varepsilon$ cannot be reduced excessively. This is a necessary condition of our model to satisfy current astrophysical constraints, as evidenced by the distribution of white dots in the figures.

\begin{figure*}[t!]
    \centering
    \includegraphics[width=0.49\linewidth]{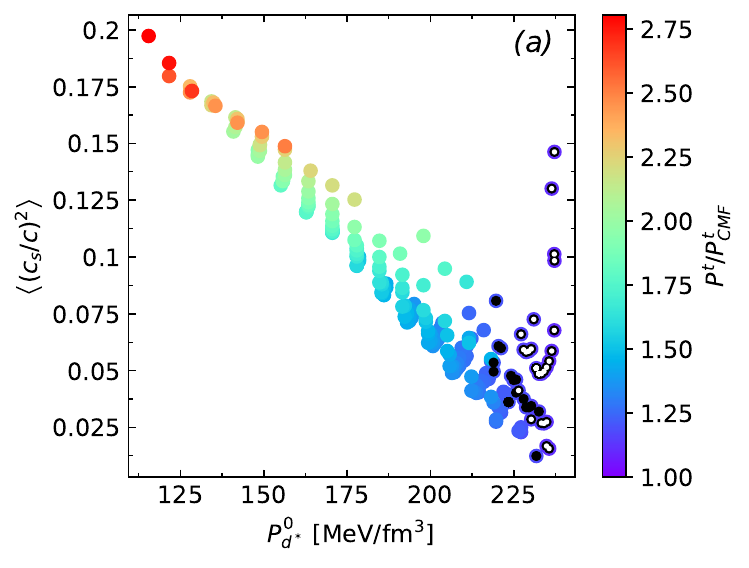}
    \includegraphics[width=0.49\linewidth]{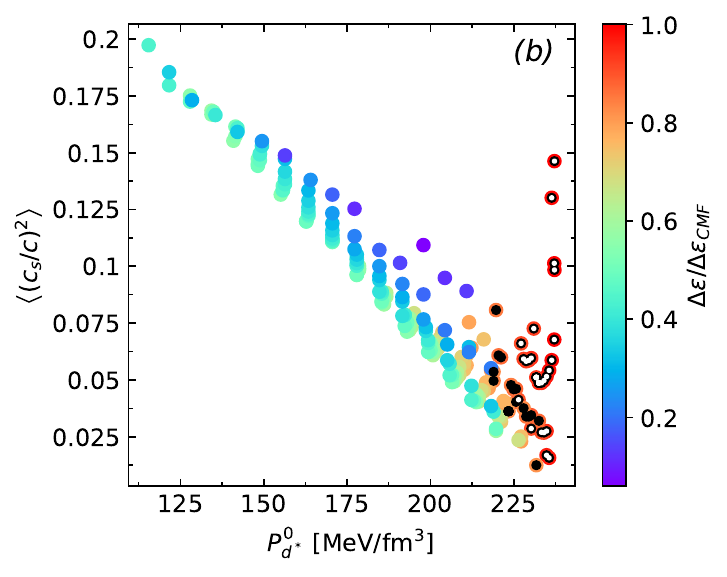}
    \includegraphics[width=0.49\linewidth]{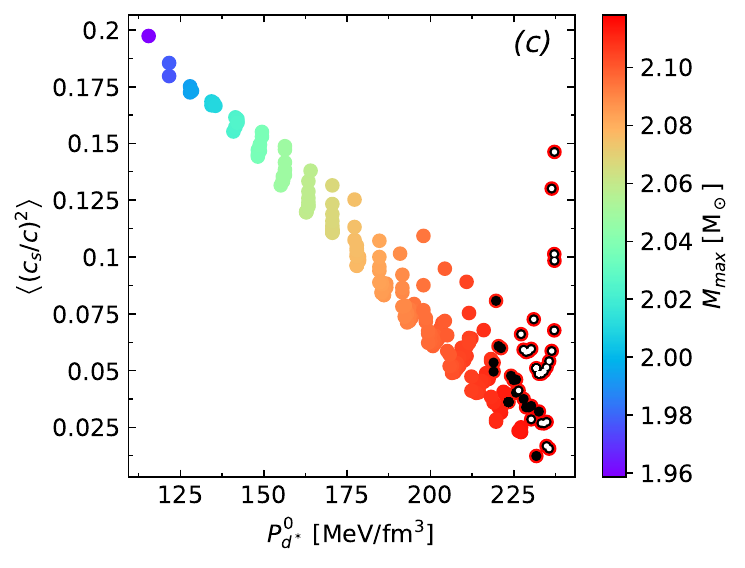}
    \caption{Three-dimensional plots showing the pressure of \dyb appearance, \pdyb, and the mean value of the squared speed of sound, \cs, in the region bounded by $P^0_{d^*}$ and $P^{\rm{t}}$, with relevant physical quantities. Panel (a): Pressure at phase transition normalized by the phase transition pressure for the CMF model without \dyb. Panel (b): Energy density jump, normalized by the CMF energy density jump  without \dyb. Panel (c): Maximum mass.}
    \label{fig:p-vs}
\end{figure*}

To analyze our results from a different perspective, we adopt two quantities previously discussed, \pdyb and \cs, as axes for our next figure, Fig.~\ref{fig:p-vs}. This approach aims to make our analysis as model-independent as possible, minimizing the dependence on the specific microphysical parameters of our model, \gsd and \gwd. We select \pdyb and \cs for this purpose, as an analogy to the CSS parametric quark model~\footnote{The CSS parametric quark model introduced in \citet{alford:2013gcf} has three independent parameters to characterize the phase transition and the quark matter EoS in a purely morphological way, $P^{\rm{t}}$, $c_s^2$ and $\Delta \varepsilon$; the appearance of \dyb particles does not involve a phase transition, there is no energy density gap.
Therefore we introduce two parameters: the appearance pressure of \dyb, \pdyb, and a proxy of the slope of the EoS when de \dyb are present, \cs.}.
In the following, we present the results in the \pdyb-\cs plane in Fig.~\ref{fig:p-vs}, with color maps representing $P^{\rm{t}}/P^{\rm{t}}_{\rm{CMF}}$, $\Delta \varepsilon /\Delta \varepsilon_{\rm{CMF}}$, and $M_{\rm{max}}$, respectively.

\begin{figure*}[t!]
    \centering
    \includegraphics[width=0.95\linewidth]{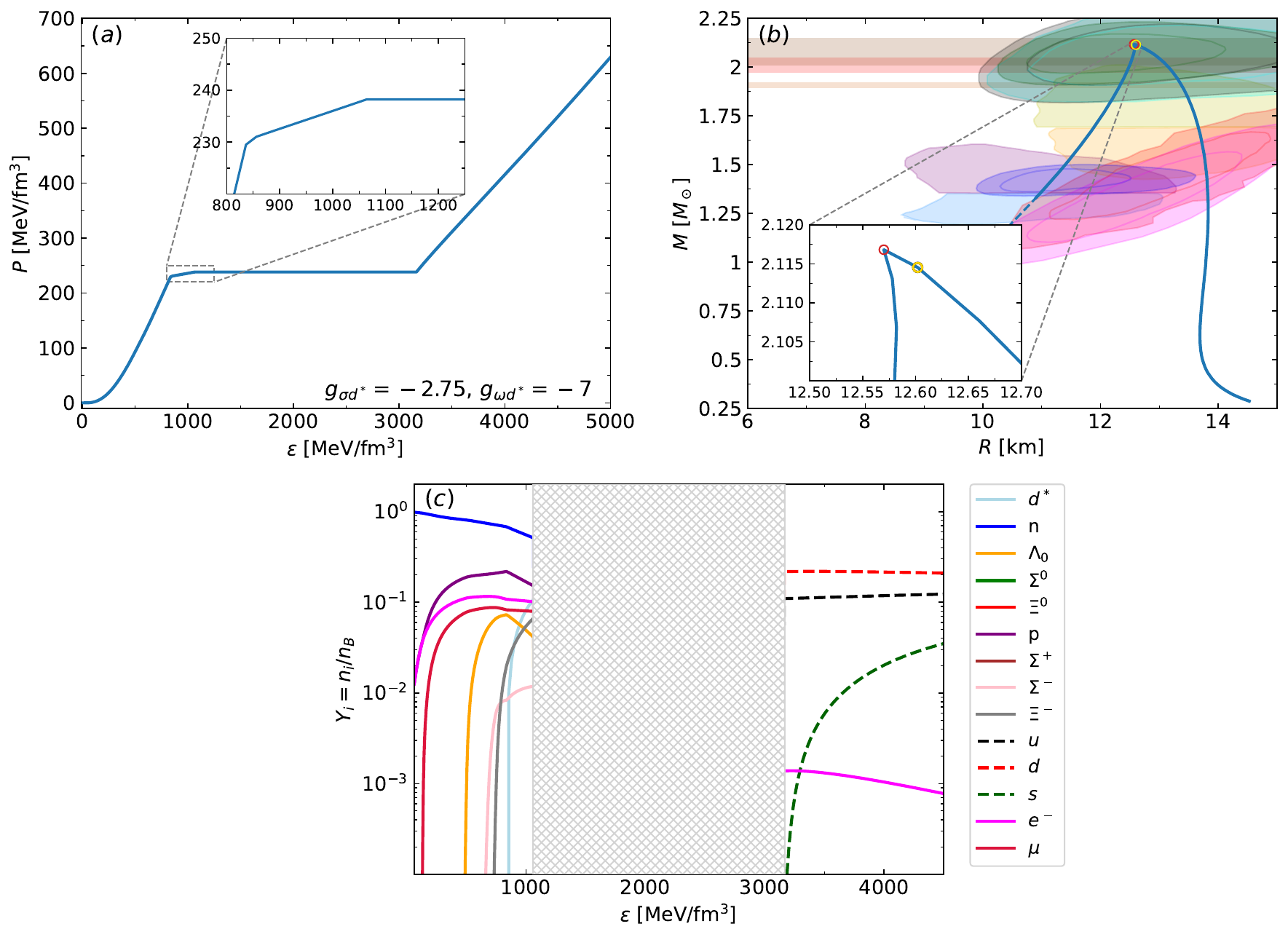}
    \caption{
    Results for a representative set of parameters, \gsd$=-2.75$ and \gwd$=-7$, corresponding to stable HSs containing \dyb that satisfy all astrophysical constraints for NSs presented in this work (sets marked with white dots in Figs.~\ref{fig:omega-sigma}, ~\ref{fig:p-vs}). Panel (a): EoS relationship. The enlarged region shows the detail of the \dyb appearance at \pdyb, where the ascending hadron curve changes its behavior, and the transition pressure marking quark deconfinement, flattening the curve at constant pressure $P^{\rm{t}}$. Note that \cs is calculated in this region, providing a measure of the \dyb relevance in the phase transition. Panel (b): Mass-radius diagram. The enlarged region shows the detail of the threshold of \dyb appearance (yellow circle) and quark deconfinement (red circle). Panel (c): Particle fractions as a function of energy density. The gray region represents the energy density jump at the transition pressure from hadron to quark matter.}
\label{fig:mosaic-selected}
\end{figure*}

In panel (a) of  Fig.~\ref{fig:p-vs}, we present the phase transition pressure, $P^{\rm{t}}$, and the energy density jump, $\Delta \varepsilon$, normalized to the corresponding values of the model without \dyb, $P^{\rm{t}}_{\rm{CMF}} = 237.23$~MeV/fm$^3$ and \mbox{$\Delta\varepsilon_{\rm{CMF}} = 2303.58 $~MeV/fm$^3$.} The black and white dot subsets have the same meaning as in previous figures. It can be seen that \pdyb and \cs have a strong correlation up to a pressure value corresponding to \mbox{\pdyb$\lesssim P^{\rm{t}}_{\rm{CMF}}$}, where the correlation breaks down. The significant dispersion in \cs in this narrow pressure range is associated with the fact that the \dyb branch is very short when \pdyb is close to $P^{\rm{t}}_{\rm{CMF}}$. Consequently, the average statistic used to construct \cs relies on very few points, leading to a large dispersion error. For hybrid EoSs with \mbox{\pdyb$>P^{\rm{t}}_{\rm{CMF}}$} no {\dyb}s are present, and thus these sets are not shown. Also, although there is a phase transition for a wide range of \pdyb and \cs, the presence of quark matter inside stable NSs (black dots) only occurs when $P^{\rm{t}}$  is slightly larger than \pdyb: as the \dyb particles tend to destabilize the NSs, quark matter must appear \textit{shortly} after the onset of hexaquarks to be present in stable stellar configurations before the maximum mass is reached. In particular, panel (a) of Fig.~\ref{fig:p-vs} shows that, in general, lower values of \pdyb or higher values of \cs correspond to monotonically higher transition pressure, $P^{\rm{t}}$. This trend has an exception in the region where \mbox{\pdyb$\lesssim P^{\rm{t}}_{\rm{CMF}}$}, where the dependence of $P^{\rm{t}}$ on \cs is lost. On the other hand, in panel (b), it can be seen that the energy density gap, $\Delta\varepsilon$, presents an inverse behavior, decreasing monotonically with increasing \cs or decreasing \pdyb; there also exists the mentioned degeneracy in \cs for \pdyb$\lesssim P^{\rm{t}}_{\rm{CMF}}$. In panel (c), we show the NS maximum mass value, $M_{\rm{max}}$, for each set, with respect to \pdyb and \cs. The figure reveals that $M_{\rm{max}}$ shows almost no dependence on \cs, but a strong correlation with \pdyb. It is important to recall  that the maximum mass value for the model without \dyb is $M_{\rm{max, CMF}}=2.118~M_\odot$. Noteworthy, the presence of \dyb particles always reduces this value, allowing the mass constraint $M_{\rm{max}}\ge2.01~M_\odot$ to be satisfied only within a limited range of \pdyb, specific for all \pdyb$\gtrsim 130$~MeV/fm$^3$.

Finally, in Fig.~\ref{fig:mosaic-selected} we present results for a specific parameter set, \gsd$=-2.75$, \gwd$=-7$, as a representative case of stable HSs that include hexaquarks and satisfy all astrophysical constraints (corresponding to the sets marked with white dots in previous figures). We show the \mbox{$P$-$\varepsilon$} relationship, the $M$-$R$ relationship, and the particle population for this set. In panel (a), it can be seen that the hadron and quark phases are separated by a first-order phase transition, characterized by an energy density gap, $\Delta \varepsilon$. The enlarged region shows the appearance of the \dyb particle just before the phase transition: when the hexaquark appears, the \mbox{$P$-$\varepsilon$} relationship exhibits a sudden drop in pressure, followed by a small increase in the energy density, leading to the phase transition and the formation of a constant pressure plateau. In particular, it is important to note that the CMF model produces large energy density jumps during the phase transition, as also observed in Refs.~\cite{Dexheimer:2010ana, Kumar:2024mna}; this feature of the model is particularly relevant in our context, as we discuss later. 

The particle population in panel (c) of Fig.~\ref{fig:mosaic-selected} correlates with panel (a), with a narrow range of hexaquarks' presence before the quark phase appearance. In the $M$-$R$ relationship (panel (b)), the traditional hadronic branch (with larger radii to satisfy GW170817 and J0437-4715 constraints) and the hybrid extended stability branch (extending beyond the maximum toward smaller radii up to the end of the continuous curve) are clearly visible. The dashed line represents unstable TOV solutions. Near the maximum mass configuration, we have included two small circles to indicate the \dyb appearance (yellow circle) and the quark phase onset (red circle). The proximity of these two dots shows the nearly immediate phase transition following the appearance of the \dyb particle. Also, the appearance of the quark phase induces a kink in the M-R curve and the occurrence of the maximum mass configuration.
Without the SSHS branch, HS configurations would be negligible or entirely absent. The existence of the (long) SSHS stability branch is crucial for satisfying the GW170817 and J0437-4715 constraints when considering stiff hadronic EoSs, such as the model and parametrization we use in this work. 
The requirement of a phase transition that produces a particularly long extended branch explains why, despite an extensive exploration of the parameter space, only a few sets satisfy all astrophysical constraints.
This situation is further aggravated by the fine-tuning needed to obtain both the presence of hexaquarks and a quark phase. Consequently, the specific set we present in detail serves as a good representative example of the white dot subset: a brief appearance of \dyb particles just before the phase transition, resulting in a long enough SSHS branch.

Regarding the length of the SSHS branch, it is closely connected to the previously  mentioned large energy density jumps in the EoS. As discussed in Refs.~\cite{lugones:2023ama, Mariani:2024cas}, there is a strong correlation between the size of the energy gap, $\Delta \varepsilon$, in the hadron-quark phase transition and the length of the SSHS branch. As the CMF model produces large values of $\Delta \varepsilon$, it is possible to obtain, for certain parameter sets, large enough SSHS branches to satisfy the small radius constraints imposed by GW170817 and J0437-4715. It is worth noting that assuming a weaker $\mu_B$ dependence in the $U_\Phi$ potential term related to the CMF deconfinement mechanism would significantly weaken the strength of the deconfinement phase transition at $T=0$ \cite{Dexheimer:2020rlp,Clevinger:2022xzl,Kumar:2023qcs} and decrease the values of $\Delta \varepsilon$. We plan to explore the role played by the hexaquark in such a scenario in a future work.

\section{Summary and conclusions} \label{sec:summary}

In this work, we have studied the interplay between the hexaquark \dyb$(2380)$ particle and quark deconfinement in HSs. We have implemented a re-parametrized version of the CMF EoS model with extended hadronic and quark phases (in terms of density) and included the \dyb in its hadronic sector. This paper inherits the development of the CMF model made in, e.g., Refs.~\cite{Dexheimer:2010ana,Kumar:2024mna}, and is a continuation of previous work based on an RMF model~\cite{Celi:2024doh}, where we studied the impact of the \dyb on neutron star stability only considering hadronic EoSs. In this work, we incorporate the possibility of a deconfinement phase transition and aim to study the effects of constructing a hybrid EoS with hexaquarks on the structure, particle population, and stability of NSs. With this in mind, we include a key ingredient into our model: the \textit{slow} hadron-quark conversion speed hypothesis, that gives rise to a SSHS extended stability branch in the mass-radius plane. This hypothesis was presented in Ref.~\cite{Pereira:2018pte} and some of its astrophysical implications were explored in Refs.~\cite{Ranea:2022bou, Ranea:2023auq, lugones:2023ama, Mariani:2024cas}. Within this scenario, we explored extensively the parameter space of our model, given by the \dyb coupling constants, \gwd and \gsd, and studied the possibility of satisfying the current astrophysical constraints through HSs that have hexaquarks in their cores.

We find, in agreement with the previous work of ~\citet{Celi:2024doh}, that once the \dyb particle appears it prevails over the hadronic phase, progressively becoming the most abundant hadron. This behavior, as it is a bosonic particle, leads to an abrupt \textit{softening} of the hadron EoS, which induces the sudden destabilization of stellar configurations. The inclusion of the hadron-quark phase transition contributes to this unstable scenario: as shown in Refs.~\cite{Seidov:1971tso, Lindblom:1998pta}, when considering phase transitions with large values of $\Delta \varepsilon$ (as the CMF model generates), the phase transition occurrence induces an immediate destabilization of stellar configurations. This double de-stability sources imply that both the \dyb  and the quark phase appearance occur near (or at) the maximum mass configuration. Beyond that point, stellar configurations can be stable under the SSHS hypothesis. 

Our \dyb parameter sweep reveals that the $2.01\, M_{\rm{max}}$ constraint for NSs complicates the situation if one aims to construct suitable hybrid EoSs that include \dyb. As our results show, these combined conditions constitute a fine-tuning problem. The narrow parameter range that remains after the astrophysical filtering could be understood in two ways: if the HS hypothesis is eventually confirmed, then the astrophysical observations could help strongly bound the hexaquark coupling constants; inversely, while quark matter existence inside compact objects remains conjectural, the \textit{fine-tuning} required to obtain such configurations could indicate that hexaquark-quark matter coexistence is strongly unlikely.

However, it is important to recall that our results have some degree of model dependence. The hadron sector of the re-parametrized CMF model could be classified as a \textit{stiff} EoS; it generates, after integrating the TOV equations, large radii for purely hadronic branches, \textit{i.e.} $R \simeq 14$ km, which is incompatible with the GW170817 NS merger event and with the recent NICER constraint for the pulsar J0437-4715. In this sense, our results are likely generalizable to any scenario where \textit{stiff} hadronic EoSs are considered. This \textit{stiff} family of hadronic EoSs is being challenged by recent NSs observations, and the SSHS branch is a proposal that could help retain their suitability while still reproducing the mentioned observations \cite{lugones:2023ama, Mariani:2024cas}. As already discussed in the previous section, the subset of explored EoSs that satisfy all the astrophysical constraints is small due to this issue, and, in case we had constructed a \textit{softer} hadron EoS, we might have satisfied the constraints with more sets, without the need for the SSHS branch.

In the specific case of the CMF EoS studied in this work, it combines the aforementioned \textit{stiffness} in the hadron sector with a large value for the energy density gap in the phase 
transition. We found that, as \textit{stiffness} makes it difficult to satisfy current small radius constraints, the high value of $\Delta \varepsilon$ helps address this issue. It has already been suggested that a large value of $\Delta \varepsilon$ favors long SSHS branches \cite{lugones:2023ama, Mariani:2024cas}, and our results seem to support this hypothesis. In this way, we are able to satisfy the challenging constraints through sets with the longest SSHS branches, which correspond to the largest values of $\Delta \varepsilon$ obtained.

Furthermore, our analysis of the $\sigma$ discontinuity, $\Delta \sigma$, along the deconfinement phase transition indicates that hexaquarks may have an influence on the hadron-quark phase transition. Although most hybrid EoSs exhibit $\Delta \sigma > 0$, consistent with the behavior observed in the CMF model without \dyb, we identify a small region of the parameter space where $\Delta \sigma < 0$. This suggests that the presence of hexaquarks not only affects the dynamics of the phase transition but may also play a role in fine-tuning the system to achieve specific, albeit unconventional, conditions, such as a continuous $\sigma$-meson field ($\Delta \sigma = 0$). While a deeper analysis of chiral symmetry restoration in the presence of \dyb is beyond the scope of this work, the behavior of the model near $\Delta \sigma = 0$ may indicate a deeper connection between hexaquarks and chiral symmetry restoration during the phase transition. However, further analysis is required to confirm such a conclusion.

Overall, our results show that the inclusion of \dyb hexaquarks in the CMF model affects the EoS and the stability of NSs, and also has an influence on the hadron-quark phase transition of HSs. This study provides new insights into the role of exotic particles in the dense matter of NSs and highlights the need for further theoretical and observational efforts to better understand the interplay between hadronic and quark degrees of freedom in these extreme astrophysical objects.

\begin{acknowledgments}
MC thanks CAPES (Brazil) for financial support under the program Move La America 2025. MC, MM, MGO and IFR-S acknowledge UNLP and CONICET (Argentina) for financial support under grants G187 and PIP 0169. This work is also supported by the U.K. STFC ST/V002570/1, ST/P003885/1. 
VD acknowledges support from the Department of Energy under grant DE-SC0024700 and by the National Science Foundation under grants MUSES OAC2103680 and NP3M PHY-2116686.
\end{acknowledgments}

\appendix

\section{Complete formulation of CMF and its new parametrization}
\label{appendix}

In the CMF model \cite{Dexheimer:2010ana}, various scalar and vector fields represent the complex dynamics of strong interactions. The scalar-isoscalar field, $\sigma$, describes the attractive interactions between baryons and between quarks. Similarly, the scalar-isoscalar $\zeta$ field captures strange quark dynamics, which are relevant in high-density matter where strange baryons and quarks can be present. Additionally, the scalar-isovector field, $\delta$, accounts for isospin asymmetry by providing a mechanism for mass splitting in isospin multiplets, which is especially relevant in systems with unequal numbers of, e.g., neutrons and protons, such as NSs. To model the repulsive forces that prevent matter from collapsing at high densities, the CMF model includes vector fields, such as the vector-isoscalar $\omega$ and vector-isovector $\rho$. These vector fields produce a short-range repulsive interaction among baryons/quarks and mesons, balancing the attractive forces provided by scalar fields. The model also introduces the strange vector-isoscalar field $\phi$, which is important in describing interactions within strange  matter. 

A unique feature of the CMF model is its inclusion of the dilaton field, $\chi$, which represents a hypothesized glueball field to mimic QCD’s trace anomaly. This field adds a layer of complexity, reflecting the effect of the gluon condensate. However, since its influence on baryon thermodynamic properties is relatively small, the CMF model often employs the “frozen glueball approximation”, setting $\chi$ to a constant vacuum value. The model is calibrated using constraints from lattice QCD, nuclear experiments, and astrophysical data, and describes the deconfinement of quarks from hadrons at high densities and temperatures \cite{Dexheimer:2015rna, Dexheimer:2010ana, Dexheimer:2018dhb, Kumar:2024mna}. It employs an order parameter, $\Phi$, inspired by the Polyakov loop, to model the hadron-quark phase transition dynamically. This allows for the exploration of the QCD critical point, where the first-order transition line ends and transitions into a crossover.

\begin{table}[t!]
\begin{tabular}{ccccccc}
\toprule 
 $g_{if}$      & $\sigma$ & $\zeta$ & $\delta$ & $\omega$ & $\phi$ & $\rho$  \\
\midrule
$N$       & $-9.83$    & 1.22    & $-2.34$    & 11.8     & 0      & 3.98   \\ 
$\Lambda$ & $-5.52$    & $-2.3$    & 0        & 7.87     & $-7.272$ & 0      \\ 
$\Sigma$  & $-4.01$    & $-4.44$   & $-6.95$    & 7.87     & $-7.272$ & 7.96   \\ 
$\Xi$     & $-1.67$    & $-7.75$   & $-4.61$    & 3.93     & $-11.23$ & 3.98   \\ 
$u$      & $-3$       & 0       & 0        & 0        & 0      & 0      \\ 
$d$      & $-3$       & 0       & 0        & 0        & 0      & 0      \\ 
$s$      & 0        & $-3$      & 0        & 0        & 0      & 0      \\ 
\bottomrule
\end{tabular}
\caption{Coupling constants $g_{if}$ of meson mean fields $i$ with fermions $f$ (nucleons, hyperons or quarks). These constants remain the same as in \citet{Kumar:2024mna}.} 
\label{table:couconstants}
\end{table}

The Lagrangian of the CMF model reads (up to a constant, subtracted to ensure that that there is no contribution in the vacuum):
\begin{align}
\mathcal{L}_{\rm CMF}&=\mathcal{L}_{\rm kin}+\mathcal{L}_{\rm 
 int}+\mathcal{L}_{\rm  scal}+\mathcal{L}_{\rm  vec}+ \mathcal{L}_{m_0}+\mathcal{L}_{\rm esb} +  \mathcal{L}_{\Phi}\,,
 \label{eq:L_CMF-app}
\end{align}
where each of the Lagrangian terms reads,
\begin{align}
\mathcal{L}_{\rm  kin} &=  \sum_{i \, \in \, \mathrm{fermions}}\bigg[\bar{\psi}_ii\gamma_\mu\partial^\mu\psi_i\bigg], \nonumber \\
  \mathcal{L}_{\rm int}&=-\sum_{i \, \in \, {\rm fermions}} \bar{\psi}_i\big[\gamma_0 \big(g_{\omega i}\omega+g_{\rho i}\rho+g_{\phi i}\phi\big) \nonumber  \\
    &+ g_{ \sigma i} \sigma + g_{ \zeta i} \zeta + g_{ \delta i} \delta \big]\psi_i\,, \nonumber \\
\mathcal{L}_{\rm  scal} &= -\frac{1}{2}k_0\chi_0^2(\sigma^2+\zeta^2+\delta^2)+k_1(\sigma^2+\zeta^2+\delta^2)^2  \nonumber \\
& +k_2\left[\frac{\sigma^4+\delta^4}{2}+\zeta^4+3 \left(\sigma\delta \right)^2\right]+k_3\chi_0\left(\sigma^2-\delta^2\right)\zeta  \nonumber\\ &-k_4\chi_0^4\nonumber+\frac{\varepsilon}{3}\chi_0^4\ln\left[\frac{\left(\sigma^2-\delta^2 \right)\zeta}{\sigma_0^2\zeta_0}\right], \nonumber \\
\mathcal{L}_{\rm  vec} &= \frac{1}{2} \left(m_\rho^2 \rho^2 +m_{\omega}^2 {\omega}^2+m_{\phi}^2 {\phi}^2\right)  \nonumber \\
&+g_4\left({\omega}^{4}+2\sqrt{2}\bigg(\frac{Z_{\phi}}{Z_{\omega}} \bigg)^{1/2} {\omega}^{3} {\phi}+3 \bigg(\frac{Z_{\phi} }{Z_{\omega}}\bigg)  {\omega}^{2} {\phi}^{2}\right.   \nonumber \\
&\left.+\sqrt{2} \bigg(\frac{Z_{\phi}}{Z_{\omega}}\bigg)^{3/2}{\omega} {\phi}^{3}+\frac{1}{4} \bigg(\frac{Z_{\phi} }{Z_{\omega}}\bigg)^2{\phi}^{4}\right)\,, \nonumber \\
\mathcal{L}_{m_0} &=  -\sum_{i \, \in \, \rm baryons}\bigg[\bar{\psi}_i m_0 \psi_i\bigg]-\sum_{i \, \in \, \rm quarks}\bigg[\bar{\psi}_i m^i_0 \psi_i\bigg]\,, \nonumber \\
\mathcal{L}_{\rm  esb} &=-\left[m_{\pi}^{2}f_{\pi}\sigma+\left(\sqrt{2}m_{K}^{2}f_{K}-\frac{1}{\sqrt{2}}m_{\pi}^{2}f_{\pi}\right)\zeta\right]  \nonumber \\
&-m^{HO}_3\sum_{i \, \in \, \mathrm{hyperons}}\bigg[\bar{\psi}_i  \left( \sqrt{2} (\sigma-\sigma_0) +  (\zeta-\zeta_0) \right)  \psi_i\bigg]\,, \nonumber \\
\mathcal{L}_{\Phi} &=  -\sum_{i \, \in \, \rm baryons}\bigg[\bar{\psi}_i  g_{\Phi B}\Phi^2  \psi_i \bigg] \nonumber   \\ 
& -\sum_{i \, \in \, \rm quarks}\bigg[ \bar{\psi}_i  g_{\Phi q}(1-\Phi)   \psi_i\bigg]-U_\Phi\,,
 \label{eq:detailed_Lag}
\end{align}
where the potential $U_\Phi$ has an explicit dependence on the baryon chemical potential, $\mu_B$, and the temperature, $T$. Its functional form is given by: 

\begin{align}
U_{\Phi}&=\left(a_0T^4+a_1\mu_B^4+a_2T^2\mu_B^2\right)\Phi^2\nonumber \\ &+a_3T_0^4\ln\left(1-6\Phi^2+8\Phi^3-3\Phi^4\right)\,.
\label{eq:U_Phi}
\end{align}

\begin{table}[t!]
\begin{tabular}{ccc}
\toprule
$k_0 = 2.37$ & $k_1 = 1.4$ & $k_2 = -5.55$ \\
$k_3 = -2.65$ & $k_4 = -0.23$ & $\varepsilon = 6/99$ \\
$f_k = 122$~MeV & $f_\pi = 93.3$~MeV & $g_4 = 43.93$ \\
$\sigma_0 = -93.46$~MeV & $\zeta_0 = -106.66$~MeV & $\chi_0 = 401.93$~MeV \\
\multicolumn{3}{c}{\begin{tabular}{c} $Z_\omega = 1.2579$ \qquad $Z_\phi = 2.1457$ \end{tabular}} \\
\bottomrule
\end{tabular}
\caption{Constants of the CMF Lagrangian~\citep{Kumar:2024mna}.}
\label{table:lagconstants}
\end{table}

\begin{table}[t!]
\begin{tabular}{c}
\toprule 
\quad $a_0 =-2.45$ \quad $a_1 = -1.1 \times 10^{-3}$ \quad $a_2 = -36.2 \times 10^{-3}$ \quad \\
\quad $a_3 = -0.396$ \quad $T_0 = 200$~MeV \quad \\
\bottomrule
\end{tabular}
\caption{Parameters of potential $U_\Phi$. Parameter $a_1$ has been changed from $a_1 = -1.81 \times 10^{-3}$ \citep{Kumar:2024mna} to $a_1 = -1.1 \times 10^{-3}$. Every other parameter remains the same than in \citet{Kumar:2024mna}.}\label{table:potconstants}
\end{table}

\begin{table} [t!]
\begin{tabular}{c}
\toprule
$m_\omega = 770.87$~MeV \, $m_\phi = 1007.76$~MeV \,  $m_\rho = 770.87$~MeV\\
$m_K = 498$~MeV \,  $m_\pi = 139$~MeV \\ 
$m_e = 0.511$~MeV \, $m_\mu = 105.7$~MeV\\
\bottomrule
\end{tabular}
\caption{Mesons and leptons masses used in this parametrization~\citep{Kumar:2024mna}.}
\label{table:masses}
\end{table}

\begin{table} [t!]
\begin{tabular}{c}
\toprule
\quad $m_0 = 150$~MeV \quad $m^{HO}_3 = 0.8061$ \quad\\
\quad $m^u_{0} = 5$~MeV \quad $m^{d}_0 = 5$~MeV \quad $m^s_{0} = 150$~MeV \quad\\
\bottomrule
\end{tabular}
\caption{The values of bare masses of fermions and fit parameter $m^{HO}_3$ in this parametrization~\citep{Kumar:2024mna}.}
\label{table:deltamasses}
\end{table}

The effective or in-medium mass of baryons and quarks in the CMF model are given by

\begin{align}
m_{i}^*&=g_{\sigma i}\sigma+g_{\zeta i}\zeta+g_{\delta i}\delta+\Delta m_i+g_{\Phi i}\Phi^2, \label{eq:emB}\\
m_{i}^*&=g_{\sigma i}\sigma+g_{\zeta i}\zeta+g_{\delta i}\delta+\Delta m_{i}+g_{\Phi i}\left(1-\Phi \right),
\label{eq:emq}
\end{align}
with 
\begin{align}
    \Delta m_N &= m_0\,,
\end{align}
for nucleons and
\begin{align}
    \Delta m_{H} &= m_0   - m^{\rm{HO}}_3 \left( \sqrt{2} \sigma_0 + \zeta_0  \right)\,,
\end{align}
for hyperons. Moreover, for quarks
\begin{align}
    \Delta m_u =\Delta m_d=m_0^u\,,  \quad \quad \Delta m_s =m_0^s\,.  
\end{align}
The effective chemical potential of baryons and quarks follows
\begin{align}
\mu_i^*=\mu_i-g_{\omega i} \omega-g_{\rho i} \rho-g_{\phi i} \phi\,.
\label{eq:emh}
\end{align}

In this framework, $\psi$ represents the fermionic field, while the couplings between fermions and meson mean fields are represented by the coupling constants, $g$’s. The parameters $k_i$’ are used to tune interactions involving scalar mesons and are calibrated accordingly. The parameter $\varepsilon$ relates to the QCD trace anomaly. Additionally, $m_K$, $m_\pi$, $f_K$, and $f_\pi$ denote the masses and decay constants of kaons and pions, key parameters that are set based on experimental data. The parameters $Z_i$ are called field redefined constants and are used to fit the vacuum masses of vector mesons. Furthermore, $m^{HO}_3$ is included to account for explicit chiral symmetry breaking and is fitted to reproduce empirical hyperon potential values, playing a key role in determining  the correct rest masses of the hyperon octet.

The coupling constants of the parametrization are shown in Table \ref{table:couconstants}. In Table \ref{table:lagconstants} the constants of the Lagrangian are presented. The constants $a$’s and the parameter $T_0$ are fitted to align with known features of the QCD phase diagram, particularly at high temperatures, as explained in Ref.~\cite{Kumar:2024mna}. The parametrization of the deconfinement potential $U_{\Phi}$ is shown in Table \ref{table:potconstants}. 

For the coupling of quarks and the $\Phi$ field, the value $g_{\Phi q} = 1000$ MeV has been considered. The value of $g_{\Phi q}$ is chosen to be sufficiently large to ensure that when $\Phi = 0$, baryon masses are low while quark masses are high, and the opposite holds when $\Phi$ is non-zero. As a result, at $T=0$, $\Phi = 0$ corresponds to a phase dominated exclusively by hadrons, while $\Phi = 1$ signifies a phase consisting solely of quarks. The coupling of baryons with the field $\Phi$ is set as $g_{\Phi B} = 3\,g_{\Phi q}$. Moreover, the value of the bare  
masses of fermions and parameter $m^{HO}_3$ can be found in Tables \ref{table:masses} and  \ref{table:deltamasses}, respectively.

\bibliography{apssamp}

\end{document}